\documentclass[prl,nofootinbib,twocolumn,superscriptaddress,preprintnumbers,floatfix]{revtex4-1}

\usepackage{placeins}
\usepackage{soul}
\usepackage[usenames,dvipsnames,svgnames]{xcolor}
\definecolor{redd}{rgb}{0.8, 0.1,0.2}
\definecolor{navy}{rgb}{0.05, 0.23,0.75}
\usepackage[colorlinks]{hyperref}
\usepackage{chngcntr}
\hypersetup{
     colorlinks   = true,
     citecolor    = navy,
	linkcolor = redd,
	urlcolor=navy,
	anchorcolor=blue
}
\usepackage{bbm}
\usepackage{amsfonts}
\usepackage{amsmath}
\usepackage{mathrsfs}

\usepackage{amsmath,amssymb,amsbsy,bm}

\usepackage{epsfig}
\usepackage{graphicx}              
\usepackage{url}
\usepackage{hyperref}
\usepackage{float}
\usepackage{pstricks}
\usepackage{color}
\usepackage{multirow}
\usepackage{lipsum}
\usepackage{enumitem}
\usepackage{ctable}
\newcolumntype{L}{>{\centering\arraybackslash}m{1.5cm}}
\def\mysection#1{{\it #1.---} }

\usepackage{enumitem}

\newcommand{\be}{\begin{equation}}
\newcommand{\ee}{\end{equation}}
\newcommand{\bea}{\begin{eqnarray}}
\newcommand{\eea}{\end{eqnarray}}
\newcommand{\beq}{\begin{eqnarray}}
\newcommand{\eeq}{\end{eqnarray}}

\newcommand{\abs}[1]{\left| #1 \right|} 
\newcommand{\msol}{{\rm M}_\odot} 

\newcommand{\aap}{AAP}
\newcommand{\mnras}{Mon. Not. Roy. Astr. Soc.}

\definecolor{cerulean}{rgb}{0., 0.62,0.7}
\newcommand{\rev}[1]{#1}

\newcommand{\ourtitle}{Constraining a Thin Dark Matter Disk with {\it Gaia} }
\newcommand{\rhoDM}{\rho_{\rm DM}}

\newcommand{\twiddle}{{\raise.17ex\hbox{$\scriptstyle\sim$}}}
\renewcommand{\vec}[1]{\mathbf{#1}}
\newcommand\numberthis{\addtocounter{equation}{1}\tag{\theequation}}

\newcommand{\es}[2] {\begin{equation} \label{#1} \begin{split} #2 \end{split} \end{equation}}

\begin{document}
\title{\ourtitle}
\author{Katelin Schutz}
\email{kschutz@berkeley.edu}
\affiliation{Berkeley Center for Theoretical Physics, University of California, Berkeley, CA 94720, USA}
\author{Tongyan Lin}
\affiliation{Berkeley Center for Theoretical Physics, University of California, Berkeley, CA 94720, USA}
\affiliation{Theoretical Physics Group, Lawrence Berkeley National Laboratory, Berkeley, CA 94720}
\affiliation{Department of Physics, University of California, San Diego, CA 92093, USA }
\author{Benjamin R. Safdi}
\affiliation{Leinweber Center for Theoretical Physics, Department of Physics, University of Michigan, Ann Arbor, MI 48109 }
\author{Chih-Liang Wu}
\affiliation{Center for Theoretical Physics, Massachusetts Institute of Technology, Cambridge, MA 02139}

\preprint{LCTP 17-03, MIT-CTP 4957}

\begin{abstract}
\noindent 
If a component of the dark matter has dissipative interactions, it could collapse to form a thin dark disk in our Galaxy that is coplanar with the baryonic disk. It has been suggested that dark disks could explain a variety of observed phenomena, including periodic comet impacts. Using the first data release from the {\it Gaia} space observatory, we search for a dark disk via its effect on stellar kinematics in the Milky Way. Our new limits disfavor the presence of a thin dark matter disk, and we present updated measurements on the total matter density in the Solar neighborhood.

\end{abstract}
\maketitle

\mysection{Introduction} The particle nature of dark matter (DM) remains a mystery in spite of its large abundance in our Universe. Moreover, some of the simplest DM models are becoming increasingly untenable. 
Taken together, the wide variety of null searches for particle DM strongly motivates taking a broader view of potential models.
Many recently-proposed models posit that DM is part of a dark sector, containing interactions or particles that lead to non-trivial dynamics on astrophysical scales~\cite{ArkaniHamed:2008qn,Kaplan:2009de,Kaplan:2009ag,Alves:2009nf,Cline:2012is,Cline:2013zca,Foot:2014mia,Cherry:2014xra, Hochberg:2014dra,Buen-Abad:2015ova,Tulin:2013teo,Tulin:2017ara,Randall:2016bqw,Fan:2013tia,Fan:2013tia,Fan:2013yva,Agrawal:2017rvu}. Meanwhile, the {\emph{Gaia}} satellite~\cite{2016A&A...595A...1G} has been observing one billion stars in the local Milky Way (MW) with high precision astrometry, which will allow for a vast improvement in our understanding of DM substructure in our Galaxy and its possible origins from dark sectors.

In this Letter, we apply the first {\emph{Gaia}} data release~\cite{2016A&A...595A...2G} to constrain the possibility that DM can dissipate energy through interactions in a dark sector.
Existing constraints imply that the entire dark sector cannot have strong self interactions, since this would lead to deviations from the predictions of cold DM that are inconsistent with cosmological observations~\cite{Cyr-Racine:2013fsa,Harvey:2015hha,Massey:2015dkw,Kahlhoefer:2015vua}. However, it is possible that only a subset of the dark sector interacts strongly or that DM interactions are only strong in low-velocity environments~\cite{Chacko:2016kgg,Fischler:2014jda,Raveri:2017jto}. 
In these cases, there is leeway in cosmological bounds and one must make use of smaller scale observables \cite{Foot:2013vna,Petraki:2014uza}. 
\rev{If the DM component can dissipate energy through emission or upscattering (see \emph{e.g.} \cite{Rosenberg:2017qia,Foot:2016wvj,CyrRacine:2012fz,Schutz:2014nka,Boddy:2016bbu,Cline:2013pca, Das:2017fyl,Buckley:2017ttd,Agrawal:2017rvu} for examples of mechanisms), then it can cool and collapse to form DM substructure. These interactions could result in a striking feature in our Galaxy: a thin DM disk (DD) \cite{Fan:2013yva,Fan:2013tia} that is coplanar with the baryonic disk. }

A thin DD may be accompanied by a range of observational signatures. For instance, DDs may be responsible for the $\sim$30 million year periodicity of comet impacts \cite{Randall:2014lxa}, the co-rotation of Andromeda's satellites \cite{Randall:2014kta,Foot:2013nea}, the point-like nature of the inner Galaxy GeV excess \cite{Agrawal:2017pnb,Lee:2015fea}, the orbital evolution of binary pulsars \cite{Caputo:2017zqh}, and the formation of massive black holes \cite{DAmico:2017lqj}, in addition to having implications for DM direct detection~\cite{McCullough:2013jma,Fan:2013bea}. Typically, a DD surface density of $\Sigma_{DD} \sim 10\, \msol$/pc$^2$ and a scale height of $h_{DD}\sim 10$~pc are required to meaningfully impact the above phenomena.\footnote{A thicker DD with $h_{DD}\gtrsim 30$~pc can cause periodic cratering \cite{Kramer:2016gqw}, however a larger surface density $\Sigma_{DD}\sim\,$15-20~$\msol$/pc$^2$ is required to be consistent with paleoclimactic constraints \cite{Shaviv:2016umn}.}

Here we present a comprehensive search for a local DD, using tracer stars as a probe of  the local gravitational potential. Specifically, we use the {\it Tycho-Gaia} Astrometric Solution (TGAS) \cite{2016A&A...595A...4L,2015A&A...574A.115M} catalog, which provides measured distances and proper motions for $\sim$2 million stars in common with the {\it Tycho-2} catalog~\cite{2000A&A...355L..27H}. 
Previous work searching for a DD with stellar kinematics used data from the {\emph{Hipparcos}} astrometric catalog \cite{Perryman:1997sa} and excluded local surface densities $\Sigma_{DD} \gtrsim 14 \, \msol$/pc$^2$ for dark disks with thickness $h_{DD} \sim10$~pc~\cite{Kramer:2016dqu}. As compared with {\it Hipparcos}, TGAS contains $\sim$20 times more stars with three dimensional positions and proper motions within a larger observed volume, which allows for a significant increase in sensitivity. Our analysis also improves on previous work by including a comprehensive set of confounding factors that were previously not all accounted for, such as uncertainties on the local density of baryonic matter and the tracer star velocity distribution.
We exclude $\Sigma_{DD} \gtrsim 6 \, \msol$/pc$^2$ for $h_{DD} \sim10$~pc, and our results put tension on the DD parameter space of interest for explaining astrophysical anomalies~\cite{Randall:2014lxa,Randall:2014kta,Foot:2013nea,Agrawal:2017pnb,Lee:2015fea,Caputo:2017zqh,DAmico:2017lqj}.

\mysection{ Vertical kinematic modeling}
We use the framework developed in Ref.~\cite{Holmberg:1998xu} (and extended in Ref.~\cite{Kramer:2016dqu}) to describe the kinematics of TGAS tracer stars in the presence of a DD. This formalism improves upon previous constraints on a DD that did not self-consistently model the profiles of the baryonic components in the presence of a DD~\cite{1989MNRAS.239..605K,1991ApJ...367L...9K,1989MNRAS.239..571K,Bovy:2013raa,McKee:2015hwa}.
These bounds typically compared the total surface density of the Galactic disk, measured from the dynamics of a tracer population above the disk, to a model of the surface density of the baryons based on extrapolating measurements from the Galactic midplane. However, the models did not include the pinching effect of the DD on the distribution of the baryons, which would lower the inferred baryon surface density for fixed midplane density.
Instead, Refs.~\cite{Holmberg:1998xu,Kramer:2016dqu} consider the dynamics close to the disk and self-consistently model the baryonic components for fixed DD surface density and scale heights. We summarize the key components below.

The phase-space distribution function of stars $f(\vec{x}, \vec{v})$ in the local MW obeys the collisionless Boltzmann equation. Assuming that the disk is axisymmetric and in equilibrium, the first non-vanishing moment of the Boltzmann equation in cylindrical coordinates is the \rev{Jeans} equation for population $A$,\vspace{-0.1cm}
  \beq \frac{1}{r \nu_A} \partial_r (r \nu_A \sigma^2_{A,rz}) + \frac{1}{\nu_A} \partial_z (\nu_A \sigma^2_{A,zz}) +\partial_z \Phi = 0 \,, \quad\label{vertical_boltzmann}\eeq 
  where $\Phi$ is the total gravitational potential, $\nu_A$ is the stellar number density, and $\sigma_{A,ij}^2$ is the velocity dispersion tensor.
The first term in Eq.~\eqref{vertical_boltzmann}, commonly known as the tilt term, can be ignored near the disk midplane where \rev{radial derivatives are much smaller than vertical ones}~\cite{Garbari:2011dh}. \rev{We assume populations are isothermal (constant $\sigma^2_{A,zz}$) near the Galactic plane~\cite{bahcall1984self}.} With these simplifying assumptions, the solution to the vertical \rev{Jeans} equation is 
$ \nu_A(z) = \nu_{A,0}\, e^{-\Phi(z)/ \sigma^2_A}$,
where we impose $\Phi(0) = 0$ and define $\sigma_A \equiv \sigma_{A,zz}$. 
For populations composed of roughly equal mass constituents (including gaseous populations), we then make the assumption that the number density and mass density are proportional, \emph{i.e.} $\rho_A(z)= \rho_{A,0} \,e^{-\Phi(z)/ \sigma_A^2}$. 

We connect the gravitational potential to the mass density of the system with the Poisson equation
\beq
\label{poisson}
 \nabla^2 \Phi = \partial_z^2 \Phi + \frac{1}{r} \partial_r( r \partial_r \Phi) = 4 \pi G \rho \,,
 \eeq
where $\rho$ is the total mass density of the system. The radial term is related to Oort's constants, with recent measurements showing $ \frac{1}{r} \partial_r( r \partial_r \Phi) =(3.4\pm 0.6) \times 10^{-3}$~M$_\odot$/pc$^3$ \cite{2017MNRAS.468L..63B}, which can be included in the analysis as a constant effective contribution to the density \cite{Silverwood:2015hxa}. 
Combining the \rev{Jeans} and Poisson equations under reflection symmetry yields an integral equation.
In the limiting case with only one population, the solution is $ \rho(z) = \rho_0\, \text{sech}^2 (\sqrt{2 \pi G \rho_0} \,z/\sigma) \label{ansatz} $. For multiple populations, the solution for the density profile and $\Phi$ must be determined numerically. 
We use an iterative solver with two steps per iteration. On the $n^\text{th}$ iteration, we compute 
\beq \Phi^{(n)}(z) = 4 \pi G \sum_A \int_0^{z} dz' \int_0^{z'} dz'' \rho^{(n)}_A (z'') \label{step1} \eeq
and update the density profile for the $A^\text{th}$ population,
\beq \rho_A^{(n+1)}(z) = \rho_{0, A}\, e^{-\Phi^{(n)}(z)/\sigma^2_A} \,.\label{step2}\eeq
Adding more gravitating populations compresses the density profile relative to the single-population case.

Near the midplane, the vertical motion is separable from the other components. For tracers indexed by $i$, the vertical component of the equilibrium Boltzmann equation is $ v_z \partial_z f_i - \partial_z \Phi \partial_{v_z} f_i = 0$, which is satisfied by
\beq
	\nu_i (z) = \nu_{i,0} \int d v_z f_{i,0}\left(\sqrt{v_z^2 + 2 \Phi(z)}\right) \,, \label{eq:tracer}\eeq where $f_{i,0}$ is the {vertical} velocity distribution at height $z=0$, normalized to unity. Given $\Phi(z)$, we can then predict the vertical number density profile for a tracer population. In our analysis we determine $f_{i,0}$ 
empirically and do not assume that our tracer population is necessarily isothermal.
    
\mysection{Mass Model} 
 \begin{table}[t]
\begin{center}
\begin{tabular}{l | c c } 
Baryonic Component & \multicolumn{1}{ c}{$\quad \rho(0)$ [M$_\odot$/pc$^3$]} & \multicolumn{1}{c }{$\quad \sigma $ [km/s] $\quad$} \\
\hline
Molecular Gas (H$_2$) & 0.0104$\,\pm\,$0.00312 & 3.7$\,\pm\,$0.2 \\
Cold Atomic Gas (H$_\text{I}$(1)) & 0.0277$\,\pm\,$0.00554 & 7.1$\,\pm\,$0.5 \\
Warm Atomic Gas (H$_\text{I}$(2)) & 0.0073$\,\pm\,$0.0007 & 22.1$\,\pm\,$2.4  \\
Hot Ionized Gas (H$_\text{II}$) & 0.0005$\,\pm\,$0.00003 & 39.0$\,\pm\,$4.0 \\
Giant Stars & 0.0006$\,\pm\,$0.00006 & 15.5$\,\pm\,$1.6 \\
$M_V <3$ & 0.0018$\,\pm\,$0.00018 & 7.5$\,\pm\,$2.0 \\
$3<M_V<4$ & 0.0018$\,\pm\,$0.00018 & 12.0$\,\pm\,$2.4 \\
$4<M_V<5$ & 0.0029$\,\pm\,$0.00029 & 18.0$\,\pm\,$1.8 \\
$5<M_V<8$ & 0.0072$\,\pm\,$0.00072 & 18.5$\,\pm\,$1.9 \\
$M_V>8$ (M Dwarfs) & 0.0216$\,\pm\,$0.0028 & 18.5$\,\pm\,$4.0 \\
White Dwarfs & 0.0056$\,\pm\,$0.001 & 20.0$\,\pm\,$5.0  \\
Brown Dwarfs & 0.0015$\,\pm\,$0.0005 & 20.0$\,\pm\,$5.0  \\
\hline 
Total & 0.0889$\,\pm\,$0.0071  & ---\\
\end{tabular}
\caption{\rev{The baryonic mass model that informs our priors}. \label{tab:massmodel}}
\end{center}
\vspace{-0.7cm}
\end{table}
\rev{In order to solve for the gravitational potential, we must have an independent model for the baryons.
In Table~\ref{tab:massmodel}, we compile some of the most up-to-date measurements of $\rho_{A,0}$ and $\sigma_A$} for the local stars and gas, primarily drawing from the results of Ref.~\cite{McKee:2015hwa} and supplementing with velocity dispersions from Refs.~\cite{Kramer:2016dew,Flynn:2006tm}. The velocity dispersions for gas components are effective dispersions that account for additional pressure terms in the Poisson-Jeans equation. \rev{ We include uncertainties (measured when available, estimated otherwise) and profile over these nuisance parameters in our analysis.}

We also include the energy density from the smooth DM halo and a possible thin DD component.
We model the bulk collisionless DM halo of the MW as a constant local density $\rhoDM$. Current measurements favor $\rhoDM \sim\,$0.01-0.02$\,\msol/$pc$^3$ at $\sim$kpc heights above the Galactic plane \cite{Read:2014qva}, though we will treat $\rhoDM$ as a nuisance parameter in our analysis. For the thin DD, we parametrize the density as \vspace{-0.2cm}
\beq
\label{DD}
\rho_{DD}(z) = \frac{\Sigma_{DD}}{ 4 h_{DD}}\, \text{sech}^2 \left( \frac{z}{ 2h_{DD} } \right)  \,,
\eeq
 with $\Sigma_{DD}$ and $h_{DD}$ the DD model parameters.    

\mysection{Tracer Stars}
For our analysis, we select TGAS stars within a cylinder about the Sun with radius $R_\text{max}=150$~pc and which extends 200 pc above and below the Galactic plane. This ensures that we are within the regime of validity for several key assumptions made above:
that the tilt term is negligible, that the various mass components have constant velocity dispersions, and that the radial and vertical motions are separable. 
Indeed, Ref.~\cite{Garbari:2011dh} showed in simulations and with data that these assumptions are satisfied within the $z \sim 200$~pc half-mass height of the disk. 
We also restrict to regions of the sky with average parallax uncertainties of 0.4~mas or less in our sample volume. In the Supplemental Material (SM) we explore the effects of different cuts and sample volumes. 

Within the spatial cuts outlined above, the TGAS sample is not complete. 
To account for this, we use the results of Ref.~\cite{2017MNRAS.470.1360B}, which compared the TGAS catalog counts to those of the \emph{Two Micron All-Sky Survey} (2MASS) catalog \cite{Skrutskie:2006wh} in order to determine the effective completeness as a function of position, color, and magnitude.
When taking the effective volume completeness into account, the tracer counts yield an optimal estimator for the true density with Poisson-distributed uncertainties \cite{2017MNRAS.470.1360B}.

Using the color cuts defined in Ref.~\cite{2013ApJSrawr}, we consider main sequence stars of spectral types A0-G4. 
Later spectral types have density profiles that are closer to constant in our volume, \rev{and thus less constraining.} In total, our sample contains 1599 A stars, 16302 F stars, and 14252 early G stars, as compared with $\sim2000$ stars that were used in the analysis of Ref.~\cite{Kramer:2016dqu}. When including a three dimensional model of dust in the selection function, Ref.~\cite{2017MNRAS.470.1360B} found that the difference in stellar density distributions is typically 1-2\% for a similar sample volume. 
We conservatively include a 3\% systematic uncertainty on the density in each $z$-bin, which also includes uncertainties in the selection function as estimated in Ref.~\cite{2017MNRAS.470.1360B}. We show the profile of our tracer stars in Fig.~\ref{fig:profile} with statistical and systematic uncertainties.

\begin{figure}[t]
\includegraphics[width = 0.49\textwidth]{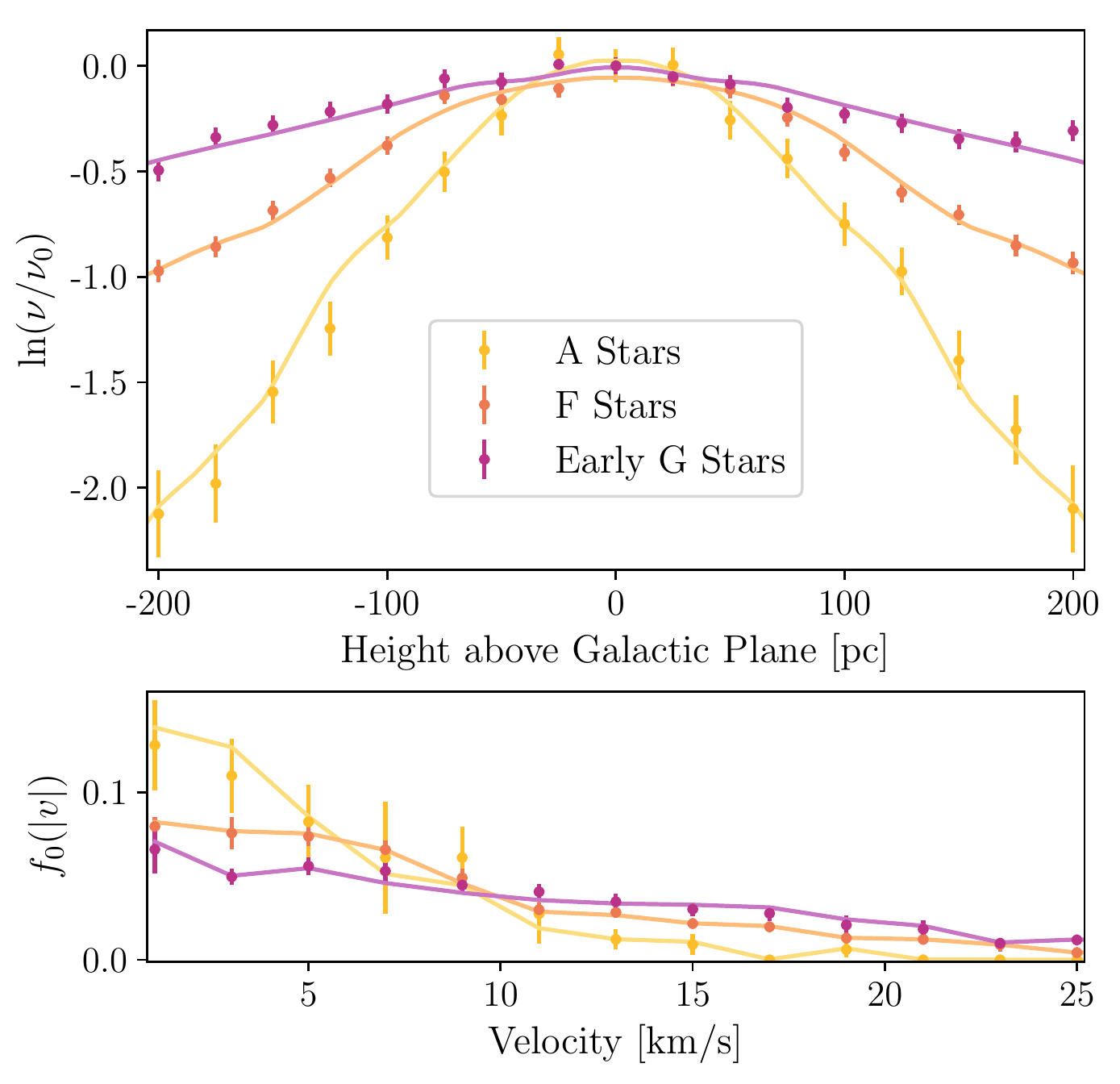}\\
 \vspace{-0.3cm}
\caption{
The measured number density profiles (top) and velocity distributions (bottom) for the tracer stars in our sample volume, subdivided by spectral type. The solid lines are the best-fit from our analysis assuming no DD.}
\label{fig:profile}
\vspace{-0.5cm}
\end{figure}

To determine the tracers' velocity distributions at the midplane, $f_{i,0}$
, we project proper and radial motions along the vertical direction as $ v_z = v_{z,0} +  (\mu_b \kappa \cos b )/\tilde{\pi} +v_r \sin b$, where $v_{z,0} $
is the vertical velocity of the Sun, $\mu_b$ is the proper motion in Galactic latitude $b$ in mas/yr, $\kappa$=4.74 is a prefactor which converts this term to units of km/s, $\tilde{\pi}$ is the parallax in mas, and $v_r$ is the radial velocity in km/s. Since the TGAS catalog does not have complete radial velocities, we perform a latitude cut $\abs{b}<5^\circ$, which geometrically ensures that radial velocities are sub-dominant in projecting for $v_z$. We then take
$v_r$ to be the mean radial velocity, $ \langle v_r \rangle = -v_{x,0} \cos l \cos b - v_{y,0} \sin l \cos b - v_{z,0} \sin b$, 
where $l$ is the Galactic longitude and where $v_{x,0} = 11.1 \pm 0.7^{\,\text{stat.}}\pm1.0^{\,\text{sys.}}$~km$/$s and $v_{y,0} = 12.24 \pm 0.47^{\,\text{stat.}}\pm2.0^{\,\text{sys.}}$~km$/$s capture the proper motion of the Sun inside the disk \cite{vzsun}. The midplane velocity distribution of our tracers is shown in the lower panel of Fig.~\ref{fig:profile} with combined statistical and systematic uncertainties, which are discussed further in the SM. 

\mysection{Out-of-equilibrium effects}
A key assumption of our analysis is that the Galactic disk is \rev{locally} in equilibrium. However, there are observations that suggest the presence of out-of-equilibrium features, such as bulk velocities, asymmetric density profiles about the Galactic plane, breathing density modes, and vertical offsets between  populations \cite{Widrow:2012wu, Williams:2013pbz,Carlin:2013eba,Yanny:2013pqi}. Such features could also be present in the DM components. While these effects are typically manifest higher above the Galactic plane than what we consider, it is still important to account for the possibility that the disk is not in equilibrium.
 
However, our tracer samples appear to obey the criteria for an equilibrium disk. When adjusting for the height of the Sun above the Galactic plane (which we find to be $-1.3\pm4.6$~pc, consistent with Refs.~\cite{2017MNRAS.470.1360B} and \cite{1997ESASP.402..721H}), we do not find any significant asymmetry in the density profile above and below the Galactic plane. We find no significant difference in the vertical velocity distribution function above and below the Galactic plane, unlike  
Ref.~\cite{Shaviv:2016yis} which claimed evidence for a contracting mode. We also find that the midplane velocity distribution function is symmetric about $v=0$ (we find the vertical Solar velocity $w_0 =6.8 \pm 0.2$ km/s, consistent with the measurement of Ref.~\cite{vzsun}) and has the expected Gaussian profile of a static isothermal population \cite{binney2011galactic}. 

Our treatment differs from the out-of-equilibrium analysis of Ref.~\cite{Kramer:2016dqu}, \rev{which evolves the \emph{observed} tracer density profile as it oscillates up and down through the spatially-fixed potential of other mass components (including a DD), while determining the error on this evolution through bootstrapping. This results in a band of possible tracer profiles that could be caused by a DD in the presence of disequilibria.} In contrast, our approach treats all the data on equal footing. Since changing $f_0(v)$ can potentially mimic the pinching effect from a DD, our analysis accounts for the possibility that pinching arises from fluctuations or systematics in $f_0(v)$. \rev{Thus, our analysis also scans over an analogous band of tracer profiles.}

We perform a final consistency check by breaking down our tracer sample into subpopulations with different velocity dispersions, which are affected differently by disequilibria due to their different mixing timescales. In the presence of out-of-equilibrium features, separate analyses of these different subpopulations could yield discrepant parameters \cite{2017MNRAS.464.3775B}. \rev{As detailed in the SM, however, we find broad agreement between the subpopulations.}

\mysection{Likelihood analysis} 
We search for evidence of a thin DD by combining the model and datasets described above with a 
likelihood function.
Here we summarize our statistical analysis, which is described in full in the SM.

The predicted $z$-distribution of stars is a function of the DD model parameters (namley the DD scale height and surface density) and nuisance parameters, which consist of:
 {\it (i)} the 12 baryonic densities in Tab.~\ref{tab:massmodel}, along with their velocity dispersions; {\it (ii)} the local DM density in the halo $\rhoDM$; {\it (iii)} the height of the Sun; {\it (iv)} the midplane stellar velocity distribution $f_0(v_j)$, where $j$ indexes the velocity bins.

\rev{The velocity distributions are given Gaussian priors} in each velocity bin with central values and widths as shown in Fig.~\ref{fig:profile}. 
The baryon densities and velocity dispersions are also given Gaussian priors with the parameters in Tab.~\ref{tab:massmodel}. The height of the Sun above the disk and local DM density are given linear priors \rev{that encompass a broad range of previous measurements,} $z_\text{sun} \in[-30, 30]$~pc and $\rho_{DM} \in [0,0.06]\,\msol$/pc$^3$~\cite{2017MNRAS.470.1360B,1997ESASP.402..721H,2007MNRAS.378..768J,Majaess:2009xc,Read:2014qva}. 
When combining stellar populations, we use a shared mass model but compute the densities of the A, F, and G stars independently and give their velocity distributions independent nuisance parameters. In analyzing all three stellar populations, we have 89 nuisance parameters.
  
\begin{figure}[t]
\includegraphics[width = 0.49\textwidth]{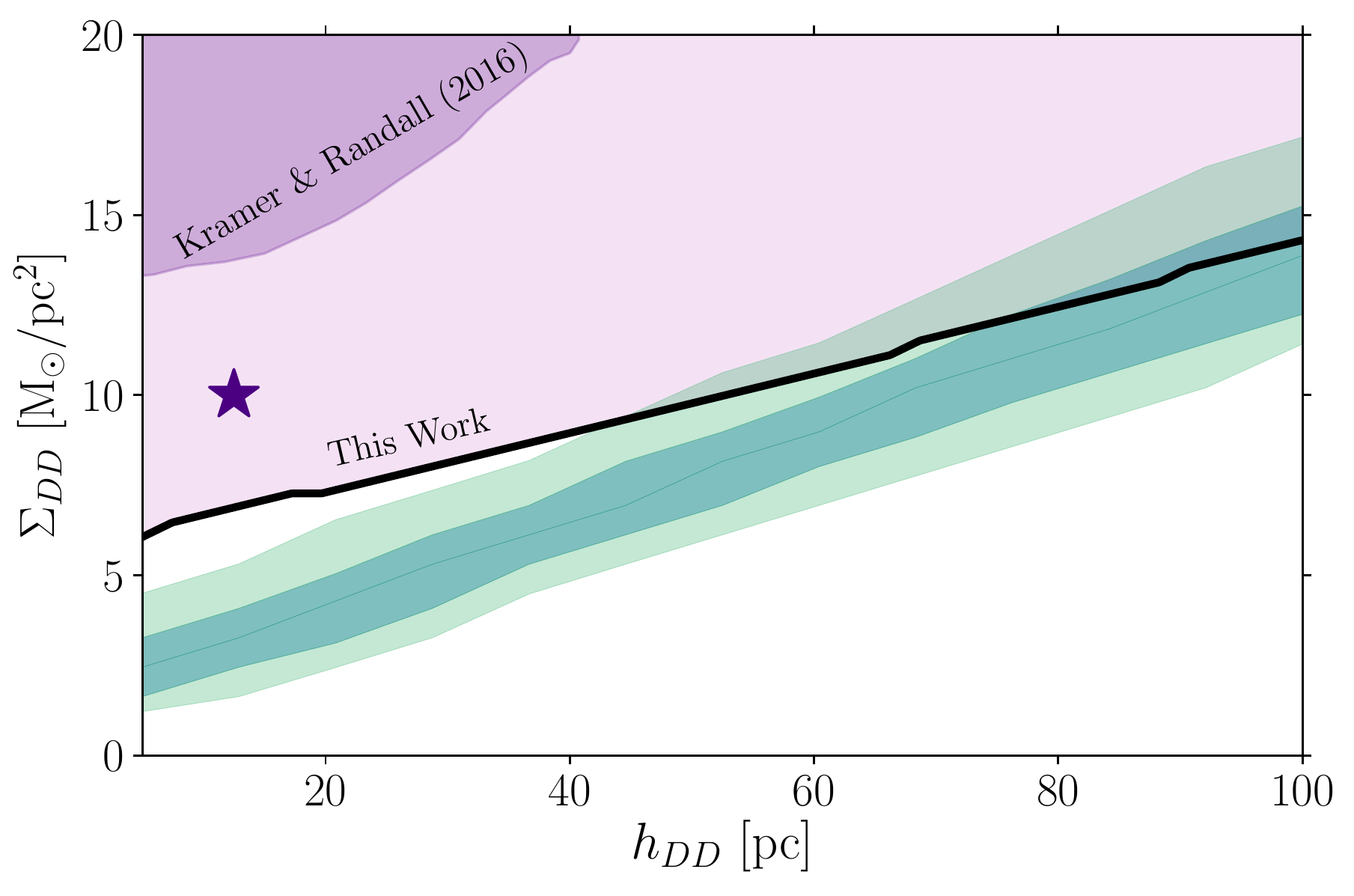}
 \vspace{-0.6cm}
\caption{The 95\% constraint on the DD surface density $\Sigma_{DD}$ as a function of the scale height $h_{DD}$, as found in this work and in Ref.~\cite{Kramer:2016dqu} (Kramer \& Randall 2016). The star indicates fiducial DD parameters that can account for phenomena such as periodic comet impacts~\cite{Randall:2014lxa}. Also shown is a comparison of the limit to the 68\% and 95\% containment regions (in dark and light green, respectively) on the expected limit from simulated data generated under the null hypothesis of no DD. 
}
\label{fig:exclusion}
\vspace{-0.4cm}
\end{figure}

 For fixed $h_{DD}$, we compute likelihood profiles as functions of $\Sigma_{DD}$ by profiling over the nuisance parameters. From the likelihood profiles we compute the 95\% one-sided limit on $\Sigma_{DD}$, which is shown in Fig.~\ref{fig:exclusion}.
  We also compare our limit to the expectation under the null hypothesis, which is generated by analyzing multiple simulated TGAS datasets where we assume the fiducial baryonic mass model and include $\rhoDM = 0.014$ M$_\odot$/pc$^3$. We present the 68\% and 95\% containment region for the expected limits at each $h_{DD}$ value. The TGAS limit is consistent with the Monte Carlo expectations at high $h_{DD}$ but becomes weaker at low $h_{DD}$. This deviation is also manifest in the test-statistic (TS), which is defined as twice the difference in log-likelihood between the maximum-likelihood DD model and the null hypothesis. 
 We find $\text{TS} \sim 5$ at $h_{DD} \sim 5$~pc and $\Sigma_{DD} \sim 4$~M$_\odot$/pc$^2$; while this does indicate that the best-fit point has a nonzero DD density, the TS is not statistically significant. Moreover we cannot exclude the possibility that, the true evidence in favor of the DD is much lower due to the possible existence of systematic uncertainties that are not captured by our analysis. 
 
 \begin{table}[h]
\begin{center}
\begin{tabular}{c  c  c  c} 
\multicolumn{1}{c}{Component} & \multicolumn{1}{ c}{A Stars}  & \multicolumn{1}{ c}{~F Stars}  & \multicolumn{1}{ c}{~Early G Stars} \\ 
\hline 
\\[-1em]
baryons&~ $0.088^{+0.006}_{-0.006}$~~& ~$0.088^{+0.007}_{-0.007}$& $0.085^{+0.007}_{-0.006}$ \\
\\[-1em]
DM & $0.038^{+0.012}_{-0.015}$ & ~$0.019^{+0.012}_{-0.011}$ & $0.004^{+0.01}_{-0.004}$ 
\end{tabular}
\caption{Posteriors on the total baryonic and DM halo density at the midplane, in units of M$_\odot$/pc$^3$. 
} \label{tab:posterior} 
\vspace{-0.3cm}
\end{center}
\end{table}

 The model without the thin DD provides insight into the baryonic mass model and the local properties of the bulk DM halo. While the DD limits described above were computed in a frequentist framework, we analyze the model without a DD within a Bayesian framework for the purposes of parameter estimation and model comparison. 
The marginalized Bayesian posterior values for the total baryonic density and local DM density are given in Tab.~\ref{tab:posterior} for analyses considering the three stellar populations in isolation. 
Despite only analyzing data in a small sample volume, we find mild evidence in favor of halo DM: for the model with halo DM compared to that without, the Bayes factors are $\sim$8.4 and $1.9$ using the A and F stars, respectively, while for early G stars the Bayes factor $\sim$0.4 is inconclusive.

\mysection{Discussion}
The results of our analysis, shown in Fig.~\ref{fig:exclusion}, strongly constrain the presence of a  DD massive enough to account for phenomena such as periodic comet impacts. 
If we assume that the DD radial profile is identical to that of the baryonic disk \rev{(which need not be the case) then we can set a limit on the fraction of DM with strong dissipations. Taking the baryon scale radius} $R_s = 2.15$~kpc \cite{Bovy:2013raa}, the Galactocentric radius of the Earth to be 8.3~kpc \cite{2017ApJ...837...30G} and the MW halo mass to be $10^{12}\msol$~\cite{2017MNRAS.465...76M}, then dissipative disk DM can account for at most $\sim$1\% of DM in the MW, for $h_{DD} \lesssim 20$ pc. \rev{Previous analyses which made this assumption found that up to $\sim$5\% of the DM in the MW could be in the disk}~\cite{Fan:2013tia}. DDs that are marginally allowed by our analysis are not necessarily stable as per Toomre's criterion \cite{toomre1964gravitational,McKee:2015hwa,Shaviv:2016umn}, \rev{although this depends on the collisional properties of the DM}~\cite{quirk1972gas} \rev{and on the presence of other disk components}~\cite{Rafikov:2000pp,Jog:2013zra}. 

For the purpose of comparing our results with previous limits, the time-dependent analysis of Ref.~\cite{Kramer:2016dqu} is the most similar to this work: \rev{although obtained in different ways, both analyses search over a band of possible density profiles that could arise from systematic effects.} Our analysis is more conservative, in that we search over multiple nuisance parameters, such as in the baryon mass model. However, we set a more stringent limit owing to the increased statistics of {\it Gaia} over {\it Hipparcos}. 

Our analysis was limited by uncertainties that can be better understood with the second {\it Gaia} data release (DR2), which will have more proper motions, spectra for measuring line-of-sight velocities, and reduced measurement errors. The improved data will reduce the systematic uncertainty on $f_0(v)$ and \rev{the line-of-sight motions will (for the first time) allow for crucial} checks on isothermality and any coupling between radial and vertical motions. Thus, the lack of evidence for local out-of-equilibrium features in DR1 can be validated with DR2.

\mysection{Acknowledgements}
We thank Ana Bonaca, Jo Bovy, Eric Kramer, Lina Necib, Mariangela Lisanti, Adrian Liu, Christopher McKee, Lisa Randall, Eddie Schlafly, Alexandra Shelest and Tracy Slatyer for useful conversations pertaining to this work. \rev{We also thank the anonymous referees.} We acknowledge the importance of equity and inclusion in this work and are committed to advancing such principles in our scientific communities. KS is supported by a National Science Foundation Graduate Research Fellowship and a Hertz Foundation Fellowship. TL is supported in part by the DOE under contract DE-AC02-05CH11231 and by NSF grant PHY-1316783. CW is partially supported by the U.S. Department of Energy under grant Contract Numbers DE-SC00012567 and DE-SC0013999 and partially supported by the Taiwan Top University Strategic Alliance (TUSA) Fellowship. This work was performed in part at Aspen Center for Physics, which is supported by NSF grant PHY-1607611.

This work has made use of data from the European Space Agency (ESA)
mission {\it Gaia} (\url{https://www.cosmos.esa.int/gaia}), processed by
the {\it Gaia} Data Processing and Analysis Consortium (DPAC,
\url{https://www.cosmos.esa.int/web/gaia/dpac/consortium}). Funding for the DPAC has been provided by national institutions, in particular the institutions participating in the {\it Gaia} Multilateral Agreement. 
This work made use of the \texttt{Multinest} nested sampling package \cite{Buchner:2014nha} through its \texttt{Python} interface \cite{2014A&A...564A.125B}, the Powell minimization algorithm \cite{powell1964efficient} implemented in \texttt{SciPy}~\cite{scipy}, 
and the \texttt{gaia\_tools} package \cite{2017MNRAS.470.1360B}.

\bibliography{gaiaDD}

\vspace{10 cm}
\clearpage
\newpage
\maketitle
\onecolumngrid
\begin{center}
\textbf{\large \ourtitle} \\ 
\vspace{0.05in}
{ \it \large Supplemental Material}\\ 
\vspace{0.05in}
{ Katelin Schutz, Tongyan Lin, Benjamin R. Safdi, and Chih-Liang Wu}
\end{center}
\counterwithin{figure}{section}
\counterwithin{table}{section}
\counterwithin{equation}{section}
\setcounter{equation}{0}
\setcounter{figure}{0}
\setcounter{table}{0}
\setcounter{section}{0}
\renewcommand{\theequation}{S\arabic{equation}}
\renewcommand{\thefigure}{S\arabic{figure}}
\renewcommand{\thetable}{S\arabic{table}}
\newcommand\ptwiddle[1]{\mathord{\mathop{#1}\limits^{\scriptscriptstyle(\sim)}}}

In this Supplemental Material, we first present the methods applied in obtaining our TGAS tracer star populations. We describe the tests performed in choosing our fiducial volume and dataset and discuss how we assign systematic errors for both density profiles and velocity distributions. We provide an extended description of our analysis, including the formalism of vertical Poisson-Jeans modeling and the full likelihood method used. The analysis is validated with mock data that mimics the TGAS data. 
We then present extended results for the main TGAS analysis presented in the text, and we further explore the dependence of the results on various systematics including choices of mass model, binning, and tracer population. 
\section{Data Selection and Systematics}
As discussed in the main Letter, our fiducial sample volume is a cylinder of radius $R_{\rm max}$ centered at the sun, with height of 200 pc above and below the Galactic plane. In selecting for the volume and the part of the sky to include in our analysis, there are several criteria which we consider below. Ideally, we would consider a sample volume as large and inclusive as possible so as to improve statistics. However, we wish to avoid biasing the sample with systematic errors that can influence our results. In this section, we demonstrate the effects of different cuts on meeting the above goals and discuss how we estimate the systematic errors on the data.

\subsection{Selection Function}
To account for the incompleteness of the TGAS data set, we implement the selection function derived in Ref.~\cite{2017MNRAS.470.1360B} and provided in the \texttt{gaia\_tools} package.\footnote{https://github.com/jobovy/gaia\_tools} In the default implementation of Ref.~\cite{2017MNRAS.470.1360B}, cuts are placed on regions of the sky in order to select for those regions with reduced measurement uncertainties. In particular, the ``good'' part of the sky is defined as those pixels with sufficient number of observations in the {\it Gaia} scan strategy, small enough variations in the scan strategy for stars in that pixel, and an ecliptic latitude cut. After these cuts, approximately 48\% of the sky is selected, with a typical mean parallax uncertainty $ < 0.45$ mas.

\begin{figure}[h]
\includegraphics[width = 0.56\textwidth]{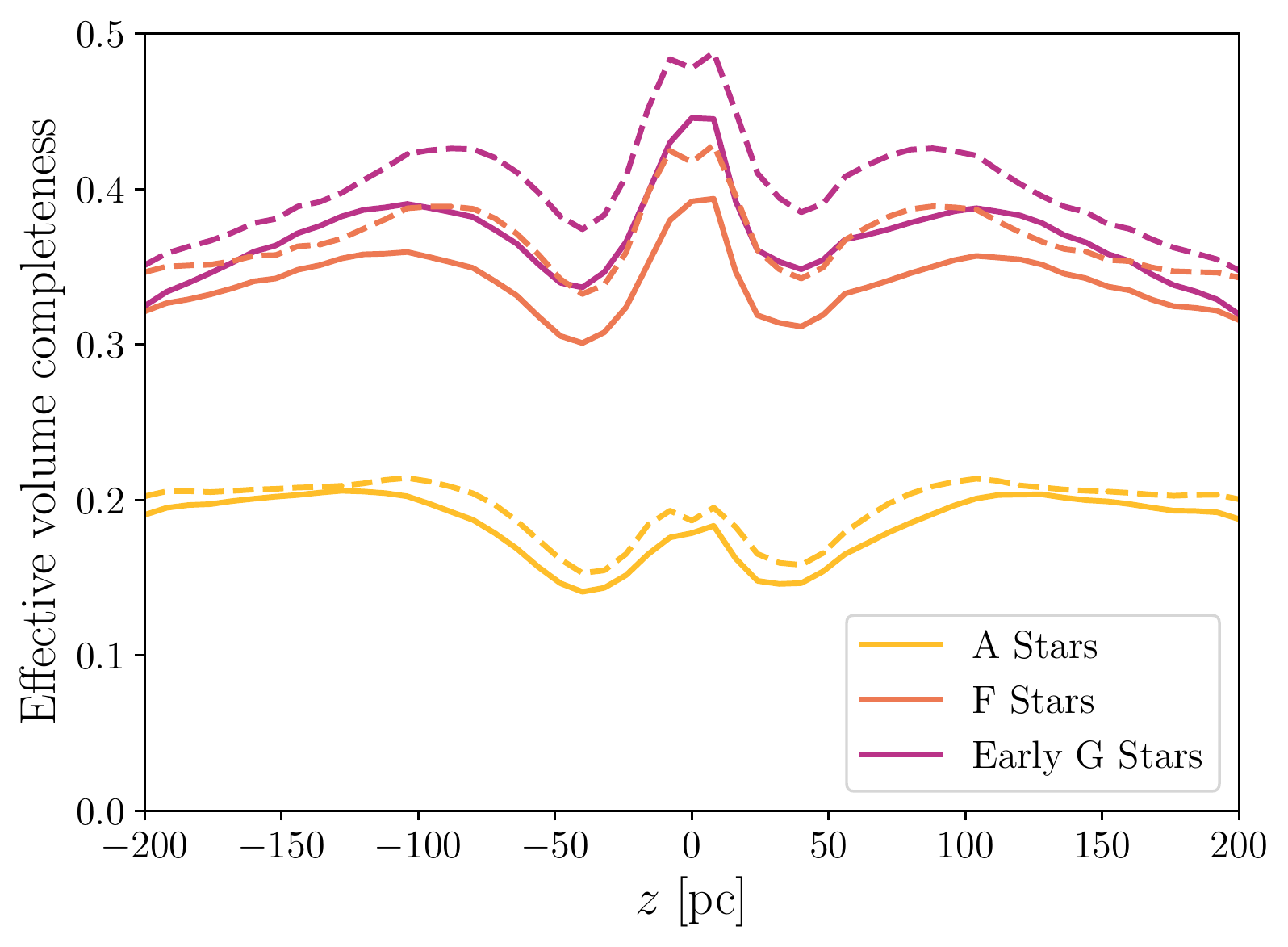}
\caption{Effectiveness completeness for our fiducial selection volume, with radial cut of $R_{\rm max} = 150$ pc. The dashed lines show the result using the default cuts in \texttt{gaia\_tools}, while the solid lines include an additional cut on regions of the sky with average statistical parallax uncertainty greater than 0.4 mas.}
\label{fig:completeness}
\end{figure}

To determine the number density of stars for a given spectral type, we then obtain the completeness for each spectral subtype ({\it e.g.} for A stars, we consider the A0, A1, ..., A9 stars individually), correct the observed number counts of stars using the effective completeness, and finally sum the subtypes. In all results for inferred density profiles shown here, we include both the systematic uncertainty on the selection function (estimated as 3\%) and statistical uncertainties computed using the observed number counts. 
In Fig.~\ref{fig:completeness}, we show the completeness for each of our tracer populations in the fiducial volume with $R_{\rm max} = 150$ pc. The selection efficiencies are fairly flat in $z$ within our sample volume, such that we are never sampling the steeply falling efficiencies at large distances. It can be seen in Figs. 3-4 of Ref.~\cite{2017MNRAS.470.1360B} that the effects of dust extinction (which is included in the 3\% systematic error budget) are also small for these distances. The dashed lines in Fig.~\ref{fig:completeness} are the result for the default selection on ``good'' parts of the sky, while the solid lines show the effect of including an additional cut on regions of the sky with average parallax uncertainty greater than 0.4~mas. In the following section, we consider the parallax uncertainty in detail.

\subsection{Parallax Uncertainties and Radial Cuts}

\begin{figure}[bt]
\includegraphics[width = 0.56\textwidth]{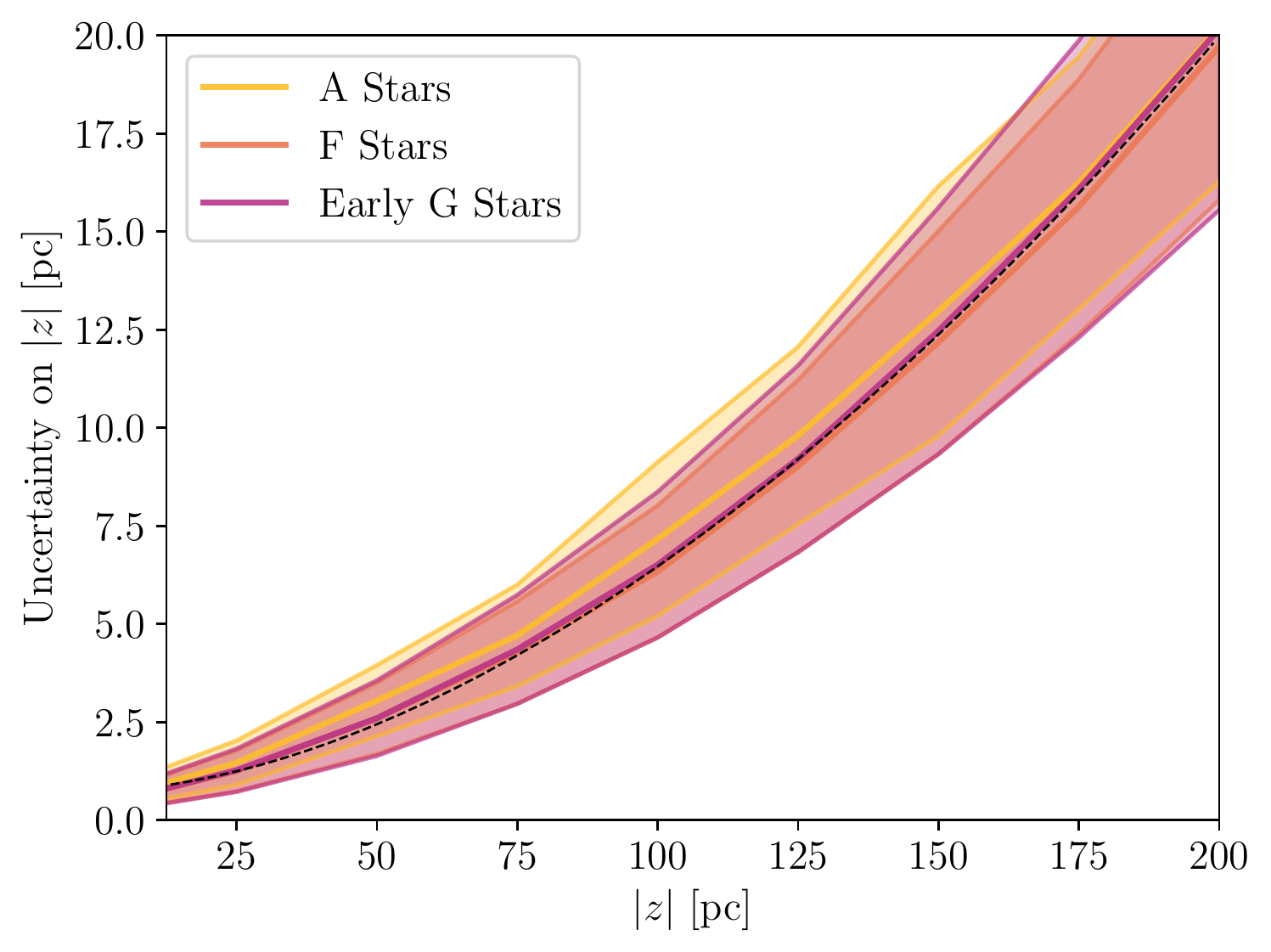}
\caption{The uncertainty in the height above the Galactic plane $z$ as a function of $z$ for a cylinder centered at Earth with radius 150~pc. The mean errors are shown as thick lines and the bands represent the spread in the error. The error in $z$ incurred from parallax uncertainties is similar between different stellar populations. The fitting function of Eq.~\eqref{eq:parallax_sigma} is shown as a black dotted line.} 
\label{fig:parallax}
\end{figure}

Perhaps the largest source of error in determining the shape of our profiles is parallax uncertainty. In order to reduce this uncertainty, we can impose a parallax cut on the data. Since cutting on individual stars could bias our sample, we instead make cuts on the \emph{mean} parallax uncertainty in a given region of the sky rather than cutting individual stars with poor parallax measurements; this optional cut is provided in the selection function of Ref.~\cite{2017MNRAS.470.1360B}. In order to retain sufficient statistics, we make a modest cut on regions of the sky with average parallax error greater than 0.4~mas, on top of the default selection in \texttt{gaia\_tools}. In addition to these statistical parallax uncertainties, there is a reported systematic parallax uncertainty on all TGAS data of 0.3~mas. In total, this translates to a $\sim10\%$ uncertainty on $z$ near the edge of our sample volume at $z=200$~pc, as can be seen in Fig.~\ref{fig:parallax}. These uncertainties can be mitigated to an extent by combining TGAS with other catalogs \cite{2017arXiv170605055A,2017arXiv170704554M}, but the issue remains that parallax uncertainties are the main contaminant in our density profiles. In our analysis, we use the fitting form 
\begin{equation}
	\sigma(z) = 0.00183 \, |z|^{1.7478} + 0.74 \, \, \text{pc} 
	\label{eq:parallax_sigma}
\end{equation} 
for the uncertainty on $z$ at different heights above the Galactic plane, as shown in Fig.~\ref{fig:parallax}.

\begin{figure}[t]
\includegraphics[width = \textwidth]{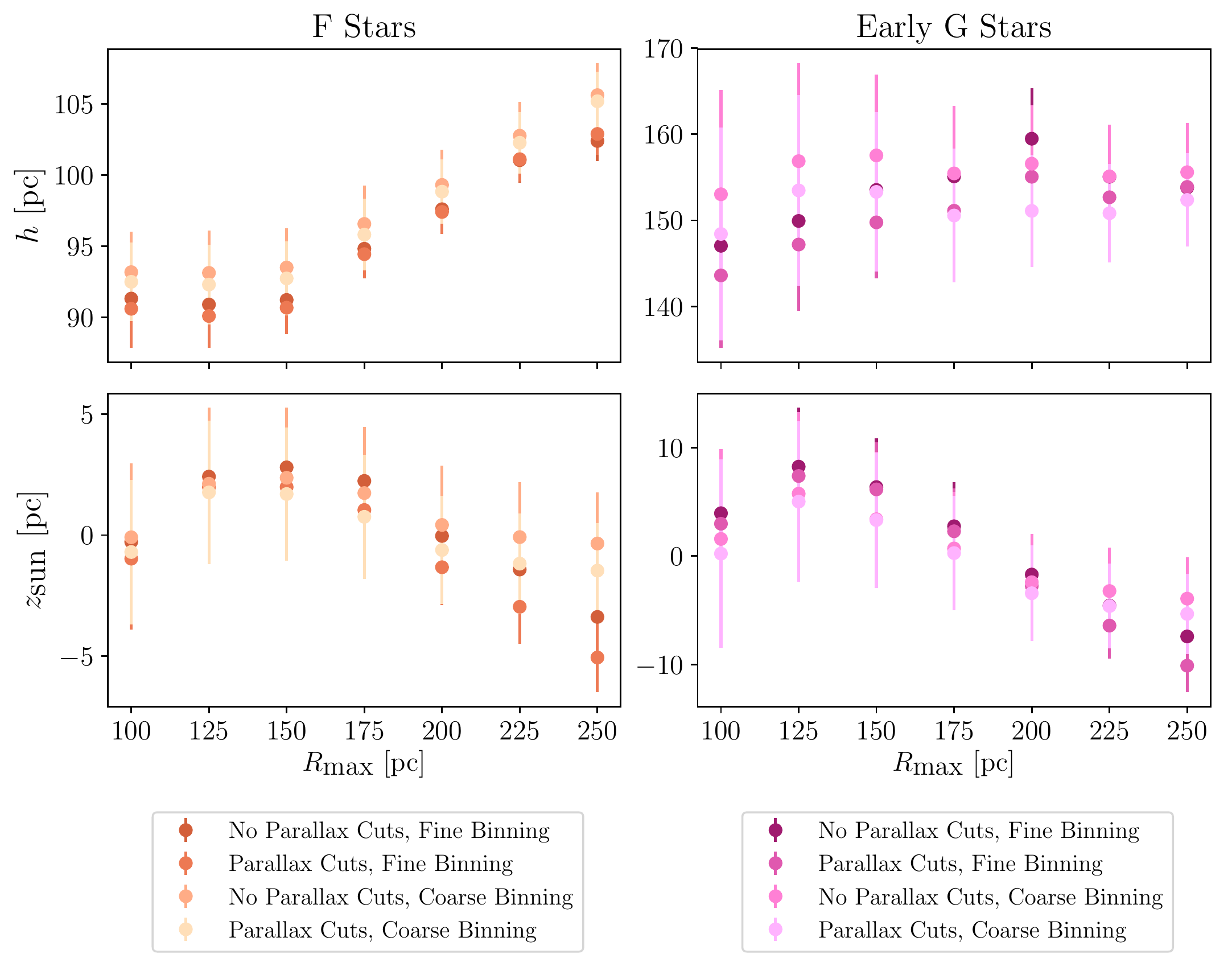}
\caption{Best fit parameters for the profile $\nu(z)=\nu_0\text{sech}^2((z+z_\text{sun})/2 h)$ of F and Early G type stars for cylindrical sample volumes with different radial cuts $R_\text{max}$. We consider data with different vertical bin sizes (8 pc and 25 pc) as well as data with and without mean parallax uncertainty cuts. Generally, placing a parallax cut or using finer binning yields a narrower inferred profile. The different choices of cuts show increasing disagreement in the best fit values for  $R_\text{max}\gtrsim150$~pc, which is suggestive of systematic effects due to parallax uncertainty.
}
\label{fig:rmax}
\end{figure}

Due to the uncertainties in parallax error, we would ideally choose bins in $z$ that are larger than the uncertainties. However, one possible effect of choosing large $z$ bins is to artificially broaden the profile. This could cause our constraints on a DD to be overly restrictive, since we are searching for a pinching effect in the profile. To check for the dependence on $z$ binning, we run our analysis on the data with both fine 8~pc bins and coarse 25~pc bins, as described in the Extended Results section of this Supplemental Material.
Parallax uncertainties also feed into the midplane vertical velocities through the projection of the proper motions onto the $z$ direction. The RMS uncertainty on midplane vertical velocities from parallax uncertainties is $\sim$0.6~km/s. Again, to check for artifacts from binning we run our analysis pipeline with fine 2~km/s velocity bins and coarse 4~km/s velocity bins in the Extended Results section of this Supplemental Material. 

In Fig.~\ref{fig:rmax}, we show the effects of the cut on the average parallax error and the choice of binning in $z$ on the tracer star density profile, as functions of the radial cut $R_\text{max}$. Fitting the density of F and early G stars to the profile  $\nu(z)=\nu_0\,\text{sech}^2((z+z_\text{sun})/2 h)$, we find systematic deviations in the scale height $h$ and height of the sun $z_\text{sun}$ depending on these choices. For both fine (8~pc) and coarse (25~pc) binning, omitting the parallax cut described in the previous section biases the profile to be broader due to larger uncertainties on $z$. The profiles are also broader with coarser binning.

\begin{figure}[t]
\includegraphics[width = 0.5\textwidth]{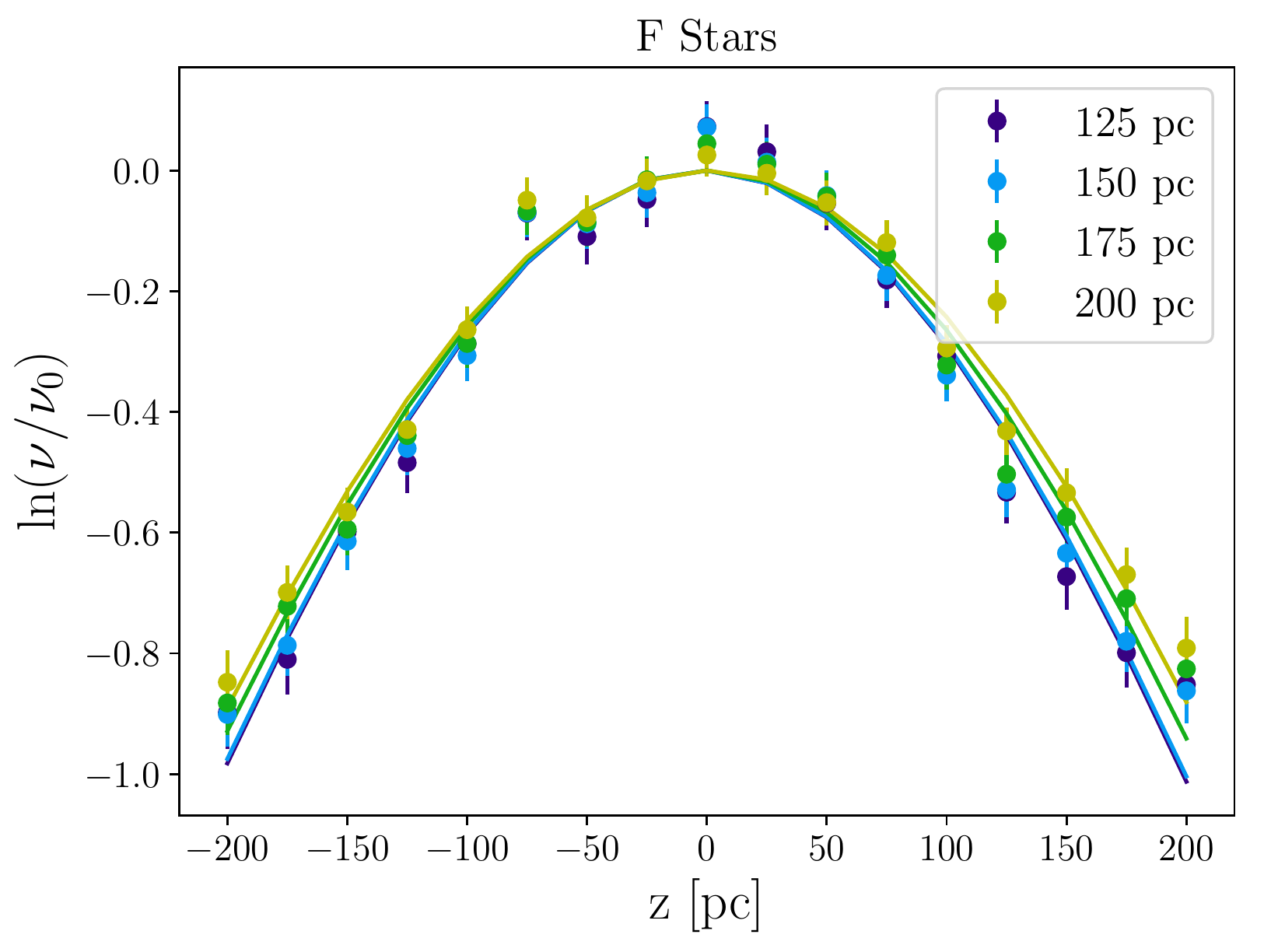}\includegraphics[width = 0.5\textwidth]{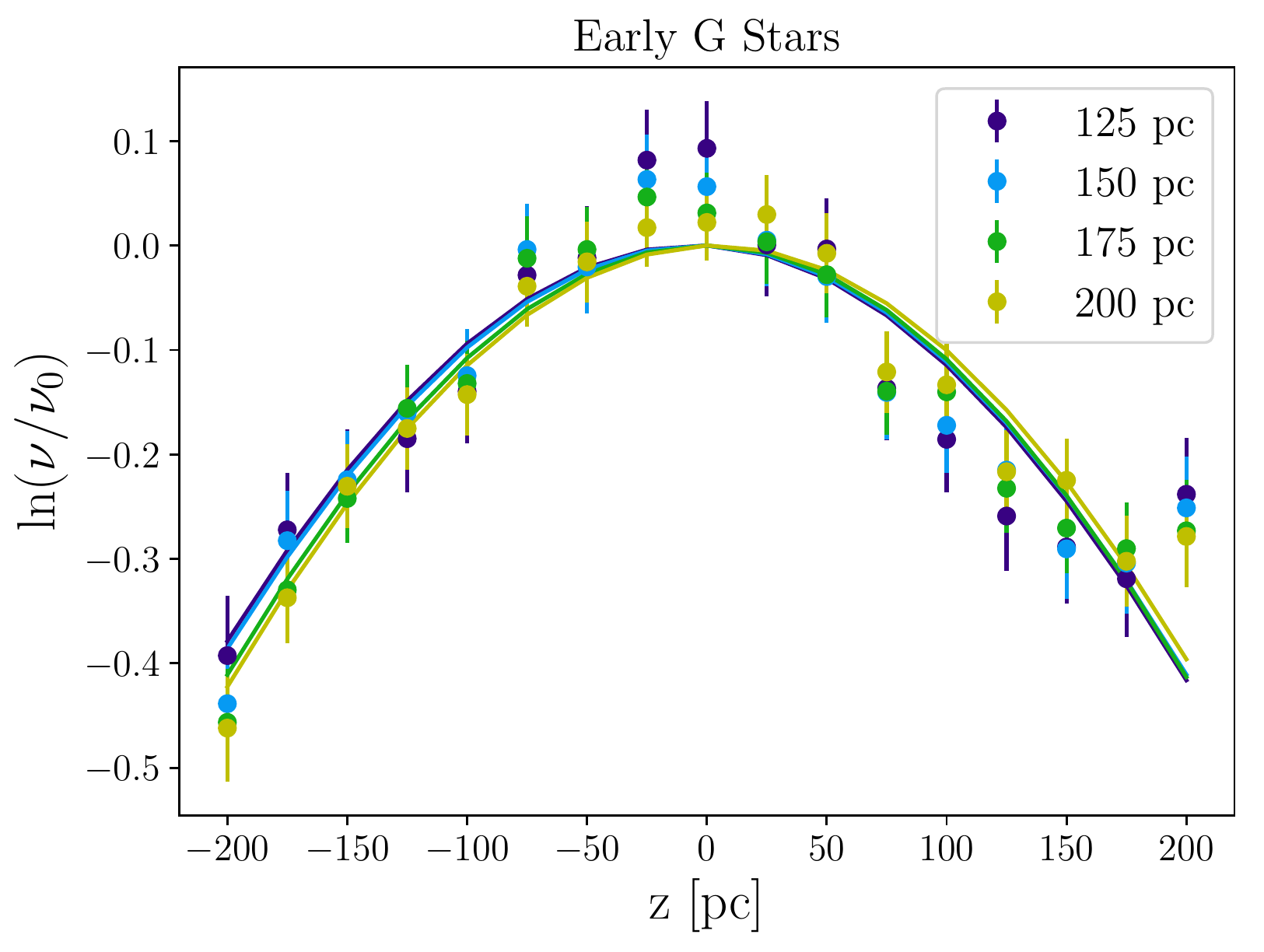}
\caption{ The density profiles $\nu(z)$ for F and early G stars in sample volumes with different values of $R_\text{max}$ and including parallax cuts. Overlaid on top of the data are the best-fit sech$^2$ profiles. As the sample volumes grow past $R_\text{max}=150$~pc, the inferred profiles broaden and shift, which we attribute to larger parallax uncertainty at the edges of the sample volume and Eddington bias. Most notably, the profiles inferred with $R_\text{max}=125$~pc and $R_\text{max}=150$~pc are in good agreement,  while the profiles inferred with $R_\text{max}=175$~pc and $R_\text{max}=200$~pc are different from the low-$R_\text{max}$ profiles and from each other. 
}
\label{fig:rcutprofiles}
\end{figure}

These differences in the density profile are mitigated by considering a smaller sample volume. We demonstrate this by showing the dependence of the fit results on the maximum radius $R_\text{max}$ of our sample volume. The most salient feature is the emergence of $R_\text{max} = $150 pc as the approximate radius beyond which the profiles start to systematically broaden more with increased radius. This is illustrated further in Fig.~\ref{fig:rcutprofiles} where a convergence of sech$^2$ profiles is seen below $R_\text{max}$=150~pc for the coarsely-binned data with parallax cuts. 

We interpret the broadening of the number density profile as being due to Eddington bias: due to the increasing parallax uncertainties with distance of the star, a larger sample volume leads to smearing of the density profiles at large $z$. In order to maximize the statistics of our analysis while reducing these systematic effects, we restrict all of the analyses in this Letter to $R_{\text max} = 150 $ pc. Since we are searching for pinching due to the presence of a DD, including data beyond $r\sim150$~pc could artificially make our constraints overly restrictive. Even with this cut, there remains non-negligible errors in the $z$-values of the stars due to parallax error, as shown in Fig.~\ref{fig:parallax}. We account for this in our analysis by applying a Gaussian kernel smoothing to our model predictions when comparing to data, as discussed further below.

\FloatBarrier
\subsection{Midplane Vertical Velocities}

In determining the vertical velocity distribution function, the primary source of uncertainty is on the radial (line of sight) velocities, which are not provided in the TGAS catalog. In our analysis, we assume that on average the stars are co-rotating with the rest of the disk and thus the mean apparent radial velocities are simply given by Earth's proper motion relative to the disk projected along the line of sight to any given star. This mean radial velocity is as large as $\sim 10$~km/s, depending on the angular position on the sky. 

Thus, in determining $f_0(|v|)$ we impose a latitude cut on stars so that the projection of the radial velocity onto the $z$ direction is small. Note that this is in contrast with other methods in the literature, such as the deconvolution approach used in Ref.~\cite{2017arXiv170403884B}. For a latitude cut of $\abs{b}<5^\circ$, the \emph{largest} contribution to the vertical velocity from the mean radial velocity is $\Delta v_z \sim 1$~km/s. This large contribution is for the special case where the line of sight to the star is roughly in the same direction as Earth's motion relative to the Galactic plane. In other parts of the sky, the radial velocity contributes a smaller uncertainty to the vertical velocity with this latitude cut. 

\begin{figure}[bt]
\includegraphics[width = 0.5\textwidth]{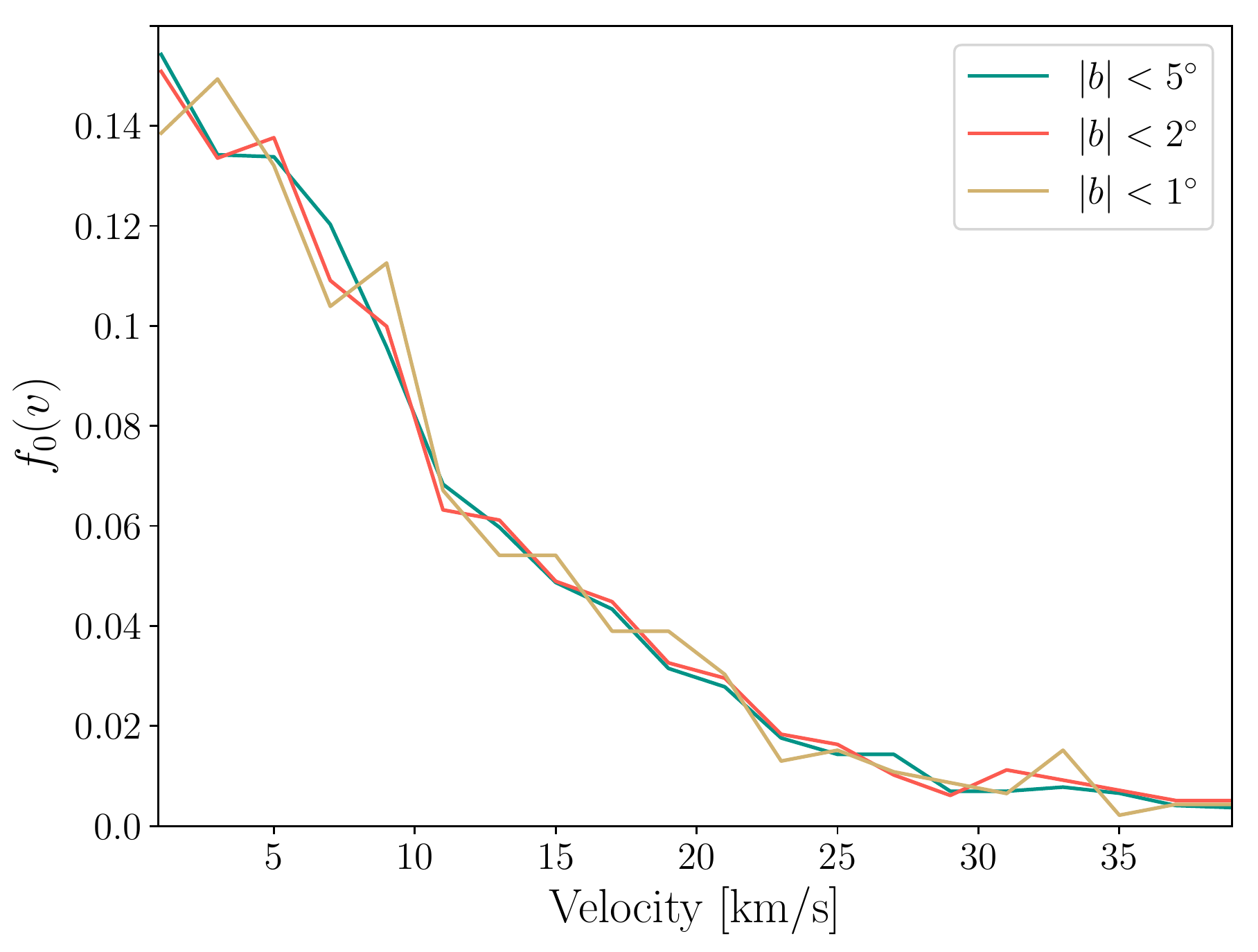}
\caption{Using the F stars as a representative case, we illustrate the effect of different latitude cuts on the inferred midplane velocity distribution $f_0(\abs{v})$. 
Tighter $b$ cuts help to mitigate the uncertainty in the radial velocity but add statistical noise. We find that within statistical uncertainties, the distribution function is not sensitive to the different latitude cuts shown here so for our analysis we use $\abs{b}<5^\circ$. }
\label{fig:bcut}
\end{figure}

\begin{figure}[t]
\includegraphics[width = 0.5\textwidth]{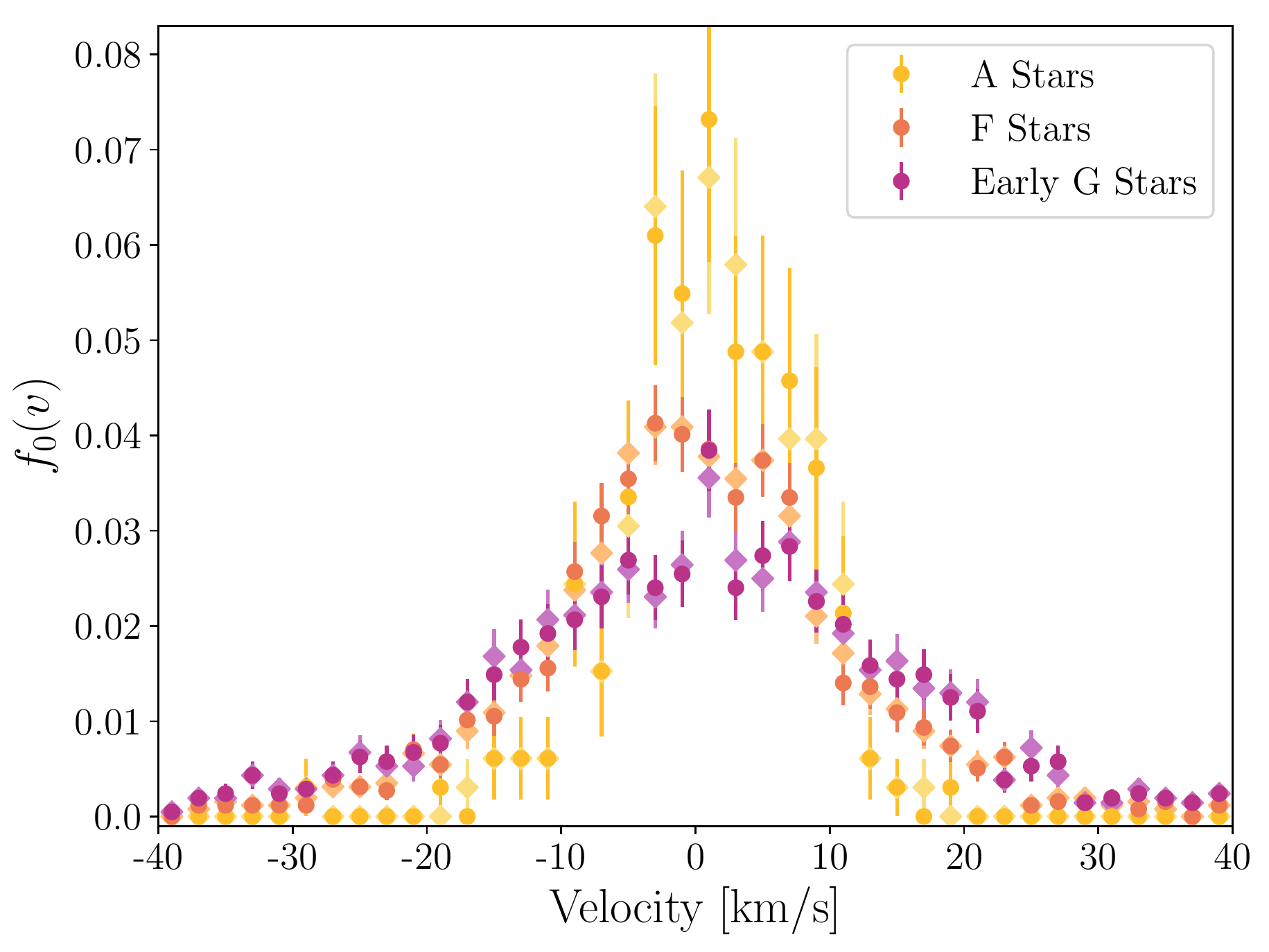}\includegraphics[width = 0.5\textwidth]{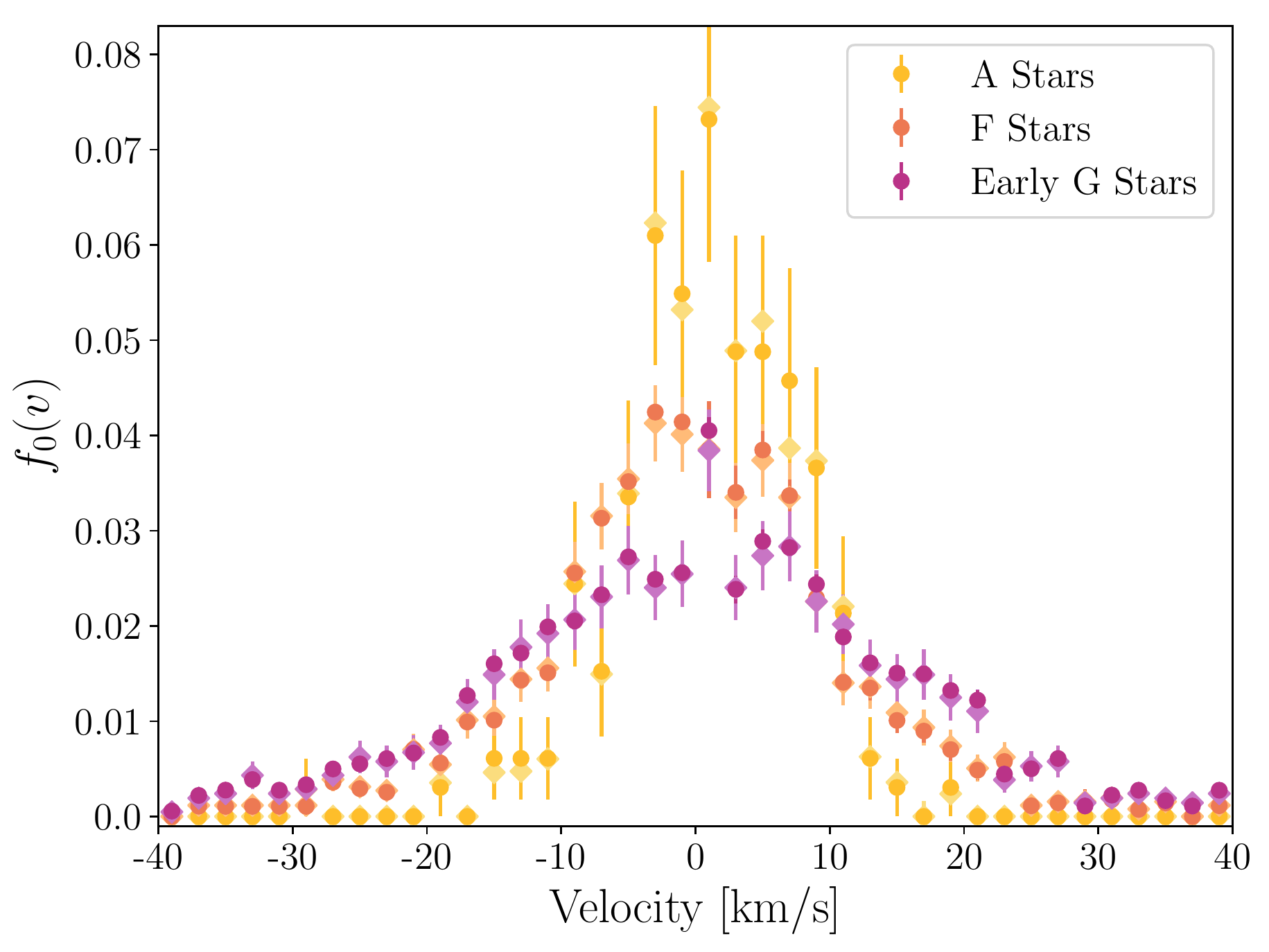}
\caption{Left: The midplane velocity distribution function for different stellar types assuming mean $v_r = \langle v_r\rangle$ (dark circles) and $v_r=0$ (light diamonds). The latitude cut $\abs{b}<5^\circ$ accounts for the smallness of the difference between these situations, which is well within the purely statistical uncertainties shown here as error bars. Right: The midplane velocity distribution function for different stellar types when appropriately weighting the stellar subtypes by their selection functions (dark circles) and when ignoring the subtype selection functions (light diamonds). Since the selection functions do not correlate with velocity but rather with position and magnitude, the two procedures are expected to be very consistent within the purely statistical uncertainties shown here as error bars. Indeed, the two procedures give very similar determinations of $f_0(v)$, with the biggest difference coming from the low-statistics A stars.
}
\label{fig:vr}
\end{figure}

The effect of the radial velocity uncertainty decreases as we make tighter latitude cuts. However, tighter cuts also reduce our statistics. In Fig.~\ref{fig:bcut} we compare the $f_0(v)$ for the F stars inferred from the data with different latitude cuts. We find good agreement between the different latitude cuts within statistical uncertainties, so we take $\abs{b}<5^\circ$ by default.
Still, the uncertainty in the radial velocities cannot be neglected; this  can be seen in the left panel of Fig.~\ref{fig:vr}, where we compare the case assuming average radial velocities to the case with radial velocities set to zero. We use the difference between these two to estimate the size of the systematic error due to radial velocities.

Also shown in Fig.~\ref{fig:vr} are the effects of the dependence of the selection function on spectral type. The selection function does not depend on velocity, although the effective completeness does vary for different spectral types (see for example Fig.~\ref{fig:completeness}). A good consistency check is to make sure that when subdividing stellar types into subtypes (\emph{e.g.} dividing A stars into types A0-A9) and weighting $f_0(v)$ accordingly by the subtype selection functions that we do not find a discrepancy with the unweighted $f_0(v)$. We find that the data are consistent between these two procedures, with the largest difference being for A stars, which have the  lowest counts. We include the difference between the $f_0(v)$ data with and without the selection function weighting as part of our estimate of the systematic uncertainty in $f_0(v)$.

\begin{figure}[t!]
\includegraphics[width = 0.5\textwidth]{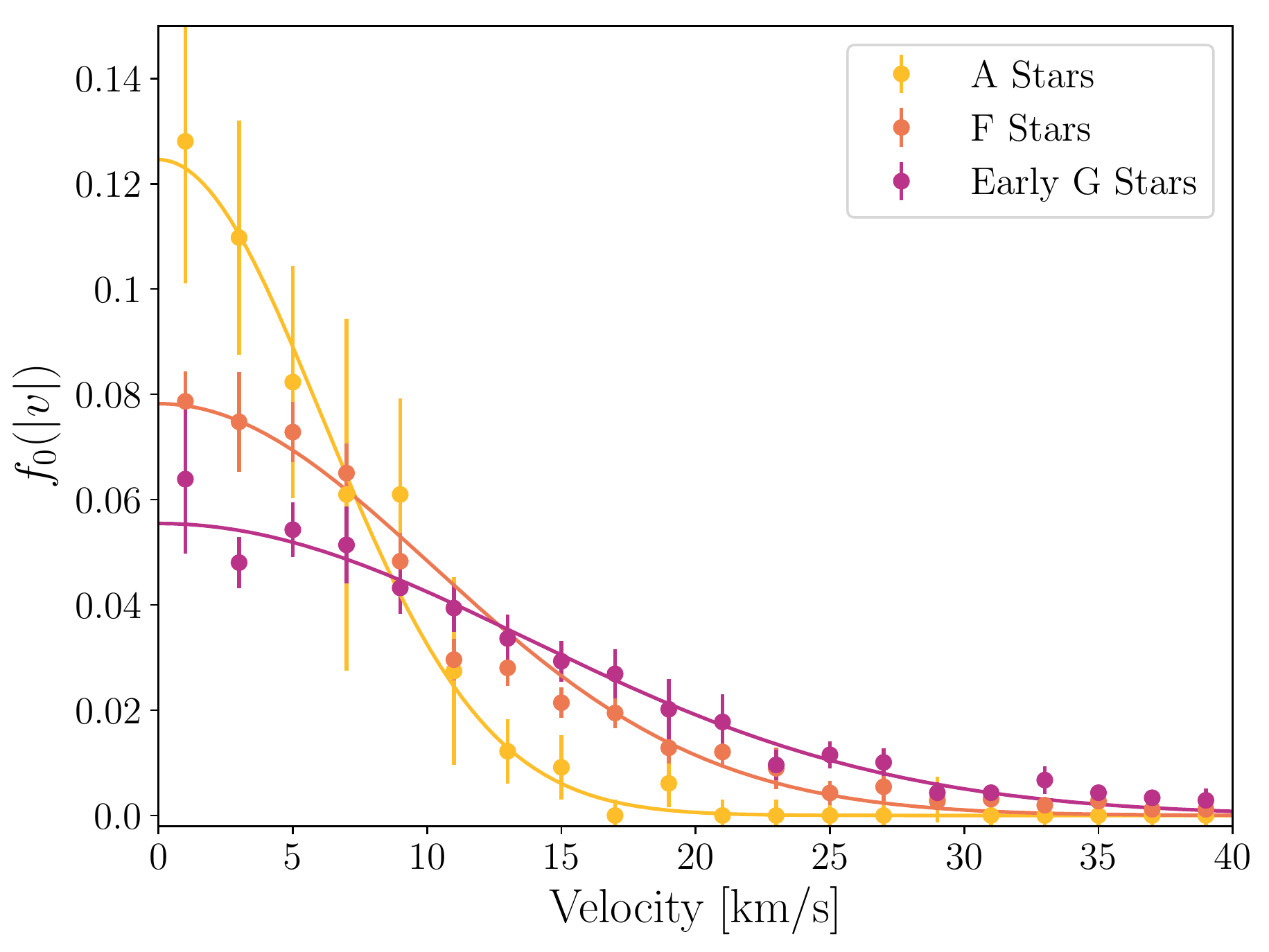}
\caption{The midplane velocity distribution function $f_0(\abs{v})$ used in our analysis. The error bars are a combination of statistical and systematic uncertainties stemming from radial velocity uncertainties, selection function weighting, and non-equilibrium-induced asymmetry in $f_0(v)$. Also shown are the best-fit Gaussian functions, which have velocity dispersions of $6.1\pm0.1$, $10.2\pm0.2$ and $13.7\pm0.1$ km/s for the A, F and G stars, respectively.
}
\label{fig:finalfv}
\end{figure}

The final contribution to the estimated systematic uncertainty comes from the asymmetry in $f_0(v)$ for positive and negative values of $v$. We find that when fitting for the vertical velocity of the Sun, which we find to be $6.8\pm 0.2$ km/s, our measurements of $f_0(v)$ for the different stellar types are well-fit by symmetric Gaussian distributions as expected in an equilibrium configuration \cite{binney2011galactic}. However, to be conservative, we take the systematic uncertainty from non-equilibrium dynamics to be the difference between the $+\abs{v}$ and $-\abs{v}$ values of $f_0(\abs{v})$. These estimated non-equilibrium systematics, in combination with the systematic uncertainties from radial velocities and selection function weighting, are included along with statistical uncertainties in Fig.~\ref{fig:finalfv} for 2~km/s velocity bins. 

\FloatBarrier

\section{Analysis procedure}
\label{sec:likelihood}

In this section, we give an extended description of our analysis procedure. We start by detailing the modeling procedure for relating the dynamics of the tracer stars to the DM and baryonic densities, following the framework described in Refs.~\cite{Holmberg:1998xu,Kramer:2016dqu}. Then, we discuss how we incorporate the model into a likelihood analysis. 

\subsection{Poisson-Jeans Theory}
In a collisionless self-gravitating system, such as a collection of stars, particles obey Liouville's theorem. In particular, for a population labeled with the upper case roman index $A$, the phase space distribution function $f_A(\vec{x}, \vec{v})$ obeys the collisionless Boltzmann equation 
\beq \frac{d\,f_A}{d \, t} = \partial_t f_A +\partial_\vec{x} f_A \cdot \vec{v} - \partial_\vec{v} f_A \cdot \partial_\vec{x} \Phi = 0 \,, \eeq
where $\Phi$ is the total gravitational potential summed over all populations and where we have dropped the explicit dependence of $f_A$ on phase space coordinates. 
Instead of working directly with $f_A$, for understanding the bulk behavior of the system it is often sufficient to describe moments of the phase space distribution, 
\begin{align} \nu_A &= \int d^3 v f_A \\
\bar{v}_{A,i}&= \frac{1}{\nu(\vec{x})} \int d^3 v \,v_i \,f_A \\
\sigma_{A, ij}^2 &= \left( \frac{1}{\nu(\vec{x})} \int d^3 v \,v_i v_j\,f_A \right) - \bar{v}_i \bar{v}_j \,,
\end{align}
such that $\nu_A$ represents the number density, $\bar{v}_{i, A}$ represents the mean velocity, and $\sigma^2_{A, ij}$ represents the velocity dispersion tensor for stars in population $A$. Note that lower case roman letters here denote spatial indices. Equipped with these definitions and assuming dynamic equilibrium (\emph{i.e.} time derivatives vanish) we can integrate moments of the Boltzmann equation in cylindrical coordinates. The first non-vanishing moment 
takes the form of Eq.~\eqref{vertical_boltzmann} 
assuming an axisymmetric disk. In general, in the Milky Way the assumption of axisymmetry (and of dropping the related time-dependence) does not hold due to the presence of the rotating spiral arms; however for the volume considered in this analysis such effects can be neglected \cite{Silverwood:2015hxa, refId0,Monari:2015uxa}. 
Next, the first term of Eq.~\eqref{vertical_boltzmann} (the tilt term) can be ignored when dealing with dynamics near the disk since \rev{radial derivatives are much smaller than vertical ones} (see \emph{e.g.} Refs.~\cite{Garbari:2011dh, Zhang:2012rsb}). However, this term can be important above the scale height of the disk \cite{Silverwood:2015hxa}. We further assume that each population is gravitationally \rev{well-equilibrated}  near the Galactic plane with constant $\sigma^2_{A, zz}$ (\emph{i.e.} that Maxwell-Boltzmann velocity statistics for an isothermal population are satisfied).

To compute the gravitational potential from the mass density of the system, we use the Poisson equation for standard Newtonian gravity given in Eq.~\eqref{poisson}. 
Since we are assuming axisymmetry, orbits are assumed to be circular so we can relate the second term to measured quantities as 
\beq \frac{1}{r}\partial^2_r (r \partial_r \Phi) =  \frac{1}{r}\partial^2_r v_c^2 = 2 ( B^2 - A^2) \,,  \eeq
where $v_c$ is the circular orbital velocity, and $A$ and $B$ are Oort's constants, defined as 
\beq A \equiv \frac{1}{2} \left(- \frac{\partial v_c}{\partial r} + \frac{v_c}{r} \right) \quad \quad B \equiv - \frac{1}{2} \left(\frac{\partial v_c }{\partial r} + \frac{v_c}{r}\right). \quad\eeq
There are a wide variety of measurements of these constants, but one of the most recent measurements comes from {\it Gaia} DR1 \cite{2017MNRAS.468L..63B}. We will adopt these values of $A = 15.3 \pm 0.4$ km/s/kpc and $B = -11.9 \pm 0.4$ km/s/kpc, which means that the radial contribution to the effective vertical density is 
\beq 
\rho_\text{eff}(z) = \rho(z) - \frac{B^2 - A^2}{2 \pi G}  = \rho(z) + (3.4\pm 0.6) \times 10^{-3}~{\rm M}_\odot/{\rm pc}^3,
\eeq 
assuming uncorrelated errors. This additional effective contribution to the density is roughly one third of previous measurements of the local DM density and can be subtracted off at the end of the analysis to obtain a measure of the physical DM density alone \cite{rhoeff}. Note that in principle Oort's constants can vary as a function of $z$; however, in the region we consider which is close to the galactic plane, this effect is smaller than the uncertainties on their measured values \cite{Bovy:2012tw}.

Combining the \rev{Jeans} and Poisson equations (again, assuming reflection symmetry) yields the integral equation
\beq \frac{\nu_A(z)}{\nu_A(0)} = \text{exp} \left(- \sum_B \frac{4 \pi G}{\sigma^2_{B, zz}} \int_0^{z} dz' \int_0^{z'} dz'' \rho_B(z'') \right) \,, \quad\quad 
 \eeq
Here we are assuming that the mass density $\rho$ is proportional to the number density $\nu$, \emph{i.e.} that while there may be some scatter in the mean mass of tracer stars, this does not have any dependence on $z$. This solution constitutes the basis of our iterative solver shown in Eqs.~\eqref{step1} and~\eqref{step2}.

Once we are equipped with a converged gravitational potential for the combined components in the Galactic disk, we can predict the vertical density profile for a given tracer population. Since we are focusing on vertical motion, we assume a form for the distribution function where the motion in the $z$ direction is separable from the other components, which is morally equivalent to dropping the tilt, azimuthal, and rotation curve terms as we have done above. Then for this $z$ component of the distribution function, the Boltzmann equation reads \beq v_z \partial_z f_i - \partial_z \Phi \partial_{v_z} f_i = 0 \,,\eeq where we have indexed tracer populations by $i$ and where again we are dropping time derivatives. 
Any function of the form $\mathcal{F} (v_z^2/2 + \Phi(z)) $ will satisfy the above differential equation. We also note that separability implies \begin{align*}& \int dv_z f_i(z, v_z) = \nu_i (z) \quad \Rightarrow \quad  f_i(z, v_z) = \nu_i(z) f_{i, z} (v_z) \,, \quad~~\numberthis\end{align*} where $f_{i, z} (v_z)$ is the velocity distribution function at some fixed height $z$, normalized to unity. One can show that in the limit of a single self-gravitating system, the velocity distribution near the midplane is Gaussian \cite{binney2011galactic}. With the above in mind, we can write 
\begin{align*}
	\nu_i(z) &= \int d v_z f_i (z, v_z)\\
	&=\int d v_z f_i\left(0, \sqrt{v_z^2 + 2 \Phi(z)} \right)\\ 
	&= \nu_i(0) \int d v_z f_{i, 0}\left(\sqrt{v_z^2 + 2 \Phi(z)}\right).\numberthis\label{eq:stracer}
\end{align*} 
Therefore, once we know the gravitational potential and the velocity distribution function for tracers at the midplane, we can solve for the tracer profile.

\subsection{Likelihood Function}
 For a single stellar population, the dataset $d$ consists of the log of the binned vertical star counts $\ln \nu_i^\text{dat}$, where $i$ labels one of the $z$ bins, and the midplane binned velocity distribution $f_j^\text{dat}$, where $j$ labels one of the velocity bins. Additionally, the dataset contains uncertainties $\sigma_{\ln \nu_i}$ and $\sigma_{f_j}$ on the number density and velocity distribution measurements, respectively, that account for both statistical and systematic uncertainties.

We use a likelihood function to fit a model $\mathcal{M}$ to the data $d$, where the model has parameters ${\bm \theta} = \{ {\bm \psi}, {\bm \zeta} \}$; the ${\bm \psi}$ are the parameters of interest and the ${\bm \zeta}$ are the nuisance parameters. Our nuisance parameters include a parameter $z_\text{sun}$ for the height of the Sun above the disk, the local DM density from the bulk halo $\rhoDM$, baryonic nuisance parameters $\rho_k$ and $\sigma_{v_k}$ for the densities of velocity dispersions of each of the $N_b$ baryonic components, indexed by $k$, listed in Tab.~\ref{tab:massmodel}, parameters $f_\nu$ that describe the overall normalization of the vertical star counts distribution for each tracer population, and nuisance parameters $f_j$ that describe the normalization of the midplane velocity distribution of the $N_t$ tracer populations in each of the $N_v$ velocity bins. Thus, the total number of nuisance parameters is $N_t + N_t N_v + 2\, N_b + 2$. Our parameters of interest ${\bm \psi}$ are the thin DD surface density $\Sigma_{DD}$ and scale height $h_{DD}$. Note that given a parameter space vector ${\bm \theta}$, we calculate the number density of stars $\nu_i( {\bm \theta} )$ in each of the $N_z$ $z$ bins though the iterative procedure described in the main Letter and the preceding section. 

As shown in Fig.~\ref{fig:parallax}, parallax uncertainties lead to a larger uncertainty on the height above the midplane at larger heights. Using this information, we can model what the true density profile would look like \emph{on average} after a simulated TGAS measurement in the presence of parallax uncertainty. We apply a Gaussian kernel to smooth the $\nu_i ({\bm \theta})$ by the TGAS parallax uncertainties, with the dispersion of the kernel varying as a function of $z$ as inferred from Eq.~\eqref{eq:parallax_sigma}, before comparing the model predictions for $\nu_i({\bm \theta})$ to data.
After taking the effects of parallax uncertainty into account in our prediction, we compare to the data through the likelihood function 
\es{likelihood:main}{
p_\nu(d | {\mathcal M}, {\bm \theta}) &= \prod_{i=1}^{N_z} {\frac{1}{\sqrt{2 \pi \sigma_{\ln \nu_i}^2} } } \exp \left[ -  { \frac{\left[ \ln (f_\nu \,  \nu_i( {\bm \theta})) - \ln \nu^\text{{dat}}_i \right]^2 }{ 2 \sigma_{\ln \nu_i}^2}} \right] \,.}
The total likelihood function for a single population in isolation is then given by the above multiplied by the appropriate prior distributions for the baryons and the stellar velocities: 
\es{likelihood}{
p(d | {\mathcal M}, {\bm \theta}) &= p_\nu(d | {\mathcal M}, {\bm \theta}) \times p_f(d | {\mathcal M}, {\bm \zeta}) \times p_b(d | {\mathcal M}, {\bm \zeta}) \,, \\
 p_f(d | {\mathcal M}, {\bm \zeta}) &= \prod_{j=1}^{N_v} {\frac{1 } {\sqrt{2 \pi \sigma_{f_j}^2}} } \exp \left[ -  { \frac{( f_{j} - f_{j}^\text{{dat}})^2 }{2 \sigma_{f_j}^2}} \right] \,; \\
  p_b(d | {\mathcal M}, {\bm \zeta}) &= \prod_{k=1}^{N_b} \left( \frac{1}{\sqrt{2 \pi \sigma_{\rho_k}^2} } \exp \left[ -  { \frac{( \rho_{k} - \rho_{k}^\text{{dat}})^2}{2 \sigma_{\rho_k}^2} }\right] \right) \left( {\frac{1} {\sqrt{2 \pi \sigma_{\sigma_{v_k}}^2}} } \exp \left[ -  { \frac{( \sigma_{v_k} - \sigma_{v_k}^\text{{dat}})^2}{2 \sigma_{\sigma_{v_k}}^2 } }\right] \right) \,.
}
Note that the prior distributions on the stellar velocities $p_f$ and baryons $p_b$ are only functions of the nuisance parameters. 
When we combine $N_t$ tracer populations, indexed by $\ell$, the likelihood function instead becomes 
\es{total_likelihood}{
p(d | {\mathcal M}, {\bm \theta}) = p_\text{baryon}(d | {\mathcal M}, {\bm \zeta}) \times \left[ \prod_{\ell=1}^{N_t} p_\nu^\ell(d | {\mathcal M}, {\bm \theta}) \times p_f^\ell(d | {\mathcal M}, {\bm \zeta}) \right]  \,.
}
Note that baryonic nuisance parameters and the nuisance parameter for the local DM density in the bulk halo are shared between all $N_t$ tracer populations, as is the nuisance parameters for the position of the Sun, while each population is given separate nuisance parameters for the binned velocities and normalization for $\nu(z)$.

Given the likelihood function, we construct likelihood profiles $\lambda( \Sigma_{DD})$ at fixed DD scale heights $h_{DD}$:
\es{LL}{
  \lambda( \Sigma_{DD} ) &= 2
    \big[ \text{max}_{{\bm\zeta }} \log p(d | {\mathcal{M}}, {\bm\theta} ) - \text{max}_{\bm \zeta, \Sigma_{DD} }   \log p(d | {\mathcal{M}}, {\bm\theta})  \big]  \,.
}
Above, the second term denotes the maximum log likelihood taken by maximizing over the nuisance parameters and $\Sigma_{DD}$, at fixed $h_{DD}$, while the first term is the maximum log likelihood at fixed $h_{DD}$ and $\Sigma_{DD}$. The likelihood profile is only strictly defined for $\Sigma_{DD}$ greater than the $\Sigma_{DD}$ of maximum likelihood. The 95\% upper limit on $\Sigma_{DD}$ is given by the value of $\Sigma_{DD}$ for which $\lambda(\Sigma_{DD}) = - 2.71$~\cite{Cowan:2010js}. The TS, on the other hand, is used to quantify the significance of a detection, and is defined analogously:
\es{TS}{
 \text{TS} &= 2
    \big[ \text{max}_{{\bm \theta }} \log p(d | {\mathcal{M}}, {\bm\theta} ) - \text{max}_{\bm \zeta}  \left. \log p(d | {\mathcal{M}}, {\bm\theta}) \right|_{\Sigma_{DD} =0 \, M_\odot / \text{pc}^{2}} \big]  \,.
}
That is, the TS is twice the log-likelihood difference between the best-fit DD model and the null model, which has $\Sigma_{DD} =0 \, M_\odot / \text{pc}^{2}$.

In the main Letter, we present both frequentist analyses, following the statistical treatment described above, and Bayesian analyses. Both types of statistical analyses use the likelihood function in Eq.~\eqref{LL}. The Bayesian analyses proceed through Bayes' theorem:
\beq 
p(\theta |\mathcal{M}, d )  = \frac{p_\nu(d|\mathcal{M}, \theta ) \, p(\theta|\mathcal{M})}{p(d|\mathcal{M})} \,.
\eeq
Above, $p(\theta | \mathcal{M})$ denotes the prior distribution; the combination of our prior and likelihood functions, $p_\nu(d|\mathcal{M}, \theta ) \, p(\theta|\mathcal{M})$, is given in Eq.~\eqref{total_likelihood}.
The posterior distribution is given by $p(\theta |\mathcal{M}, d ) $, while the Bayesian evidence (sometimes also referred to as the marginal likelihood or model likelihood) is found through the integral 
\beq p(d|\mathcal{M}) = \int d \theta \, p(d | \mathcal{M}, \theta) \, p(\theta|\mathcal{M}) \,. \eeq 
When performing parameter estimation in the Bayesian framework, we integrate the posterior distribution over all model and nuisance parameters except for our specific parameter of interest and then calculate the appropriate percentiles of the resulting one-dimensional posterior distribution. We also, in the main Letter, compare nested models using the Bayes factor. The Bayes factor in preference for a model $\mathcal{M}_A$ over a model $\mathcal{M}_B$ is given by the evidence ratio
\beq \mathcal{B}_{AB} = \frac{p(d|\mathcal{M}_A)}{p(d|\mathcal{M}_B)} \,.\eeq 

\subsection{Mass model}
A key part of our analysis is the inclusion of uncertainties on our mass model, as reported in Table~\ref{tab:massmodel}. This mass model is a composite from a variety of sources in the literature, primarily drawing from the results of Ref.~\cite{McKee:2015hwa}, where we obtained the densities and associated uncertainties in Table~\ref{tab:massmodel}. The local gas densities are determined from measurements of the column density of hydrogen in its various forms, assuming a fixed scale height. A correction factor of 1.4 is applied to account for the presence of heavier elements whose abundance relative to hydrogen is known. For the case of molecular hydrogen (H$_2$), there is a 30\% uncertainty on the density, while for the ionized hydrogen (H$_\text{II}$) the uncertainty is 5\%. The atomic hydrogen (H$_\text{I}$) is split into warm and cold components: the uncertainty on the density of the warm component is 10\% while for the cold component the uncertainty is 20\% due to optical depth corrections. 

For the giant stars and stars with $M_V<8$, we estimate the uncertainty on the density as 10\%, which is commensurate with the relative uncertainty on the \emph{total} surface density of these populations for fixed scale height in the analysis of Ref.~\cite{McKee:2015hwa}. Note that the analysis of Ref.~\cite{2017MNRAS.470.1360B} found very similar values for the density of giant stars and main sequence stars with $M_V<8$, within the margins of uncertainty reported by Ref.~\cite{McKee:2015hwa}. Since these stars are such a subdominant component of the mass model, we do not perform a separate analysis with measured densities of Ref.~\cite{2017MNRAS.470.1360B} but instead perform our analysis with those from Ref.~\cite{McKee:2015hwa}, which have more conservative uncertainties. For the M dwarfs the uncertainty on the surface density (again, for fixed scale height) is 13\%, which we translates to the same relative uncertainty for the density \cite{McKee:2015hwa}. The uncertainty on the density of white dwarfs is given explicitly in Ref.~\cite{McKee:2015hwa}, while for the brown dwarfs we assume a 35\% uncertainty on the density coming from the uncertainty in the spectral index of the brown dwarf mass function. Note that for the densities reported here, the thick stellar disk is already taken into account as described further in Ref.~\cite{McKee:2015hwa}.

For the velocity dispersions, we aggregate measurements and uncertainties from several sources. The velocity dispersions for the gas, giant stars, and stars with $M_V<4$ came from the revised estimates provided in Refs.~\cite{Kramer:2016dew,Kramer:2016dqu}, while for the other category we use velocity dispersions from Refs.~\cite{Read:2014qva,Flynn:2006tm}. In particular the updated velocity dispersions and uncertainties we quote for the gas include additional non-thermal contributions to the pressure, including the effects of magnetic fields. Where available, the uncertainties on the velocity dispersion were drawn from Refs.~\cite{Read:2014qva,Flynn:2006tm}. In some cases, the populations were categorized differently, in which case a best attempt was made to consolidate different references. For instance, Ref.~\cite{Kramer:2016dqu} estimated the velocity dispersion for the category $M_V < 3$ as being the same for the category $M_V < 2.5$, under the assumption that the contribution from stars with $2.5 < M_V < 3$ is small compared to the error on the velocity dispersion. 
For the stars with $3<M_V<8$ and the M dwarfs, we estimate the error on the velocity dispersions as coming from the variance of different measurements of the scale height aggregated in Ref.~\cite{McKee:2015hwa}, which yield similar error estimates as those reported in Refs.~\cite{Read:2014qva,Flynn:2006tm}. Finally, in other cases where the updated stellar velocity dispersions from Ref.~\cite{Kramer:2016dqu} were different from the values given in Refs.~\cite{Read:2014qva,Flynn:2006tm}, we assume that the relative size of the error on the revised values is the same as for the previous values. For instance, for the giant stars the error is assumed to be 10\% for the new value of the velocity dispersion,  as it was for the old value.

\FloatBarrier
\section{Validating our Analysis with Mock Data}

In this section we test our analysis framework on simulated TGAS data. By analyzing simulated data with an injected DD signal, we show that our analysis framework is able to appropriately reconstruct the injected parameters and importantly we demonstrate that the resulting limits to do not exclude the true DD parameters.  

\subsection{Mock Data Generation}

We create an ensemble of mock datasets in the following way. First, we assume a mass model. For the baryons, we simply take the centers of the priors for the midplane densities and velocity dispersions. For every mock dataset we set $z_\text{sun}=0$~pc and halo DM density $\rhoDM = 0.01\, \msol$/pc$^3$. We optionally include a DD in the mass model for testing the recovery of injected DD parameters. 

The velocity distributions of the three tracer populations are drawn assuming Poisson fluctuations in each velocity bin, where we assume the true $f_0(|v|)$ is given by the measured distribution of that population.
Using the mass model and the randomly drawn velocity distribution function, we then calculate the gravitational potential and density profile $\nu(z)$ for each tracer population. The calculated density profile is normalized to the observed number of tracer stars in the TGAS data over the range $z \in [-200,200]$ pc; then, using this distribution, we randomly generate both the $z$-positions of stars as well as the total number of stars assuming Poisson noise. Distance uncertainties due to parallax error are also included in this data set. For every star with distance $z_i$  we introduce a random Gaussian smearing on $z_i$ with width given by $\sigma(z_i)$ in Eq.~\ref{eq:parallax_sigma}. Finally, we assign both statistical and systematic uncertainties to the mock data: for the number densities, we assume a 3\% systematic as for the density profiles, while for the velocity distributions we assume 15\%, 5\%, and 7\% systematic errors for A, F, and Early G stars, respectively, which roughly reproduce the uncertainty levels we find in the real TGAS data.

\begin{figure}[tb]
\includegraphics[width = \textwidth]{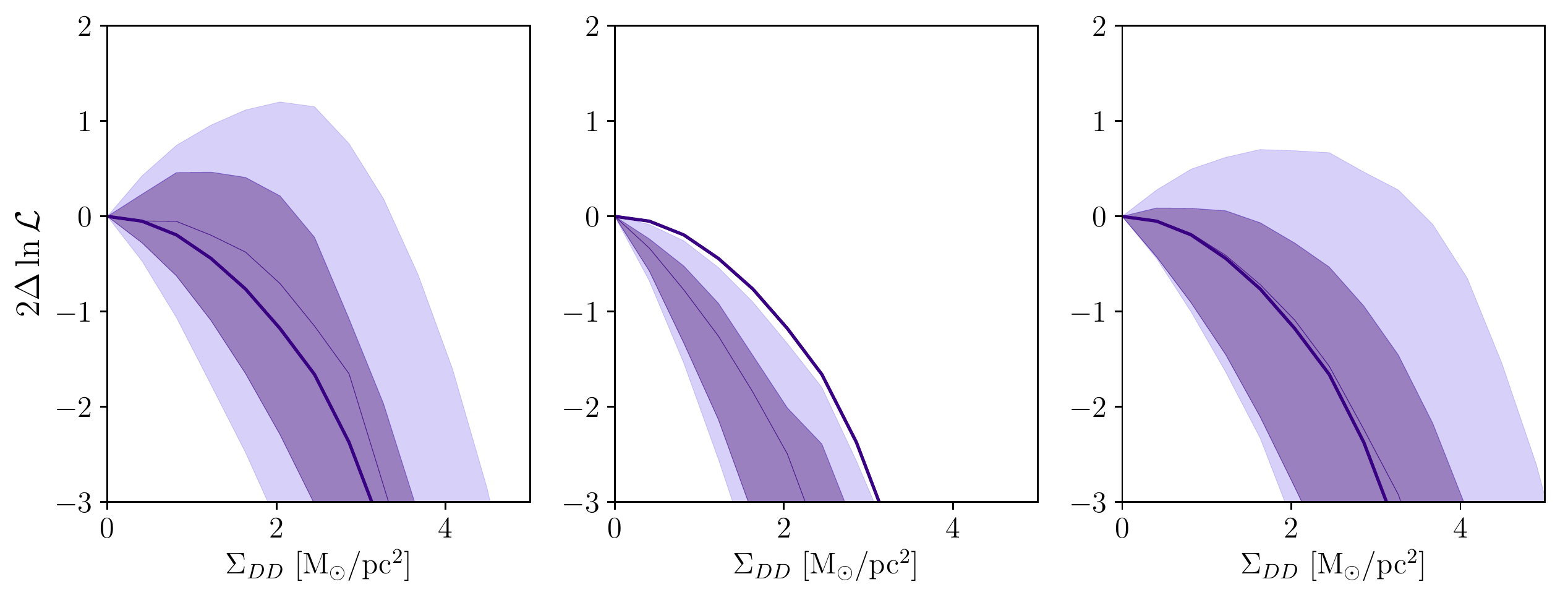}
\caption{Left: The likelihood profile when no parallax error is introduced into the mock datasets and with no Gaussian kernel density smoothing. Center: Same as the left but with parallax error introduced into the mock datasets, causing artificially tight constraints. Right: The recovered likelihood profile when parallax error is present in the mock data and when the Gaussian kernel is used in the analysis to model the effects of parallax error. The likelihood profile using the Gaussian kernel has similar statistics and yields a similar limit on a DD under the null hypothesis as the case with no parallax error. In all three panels, we take $h_{DD} = 5$ pc and no DD is injected. The dark (light) shaded regions correspond to the 68\% (95\%) containment regions, and the thick line corresponds to the results of the Asimov analysis. }
\label{fig:smear}
\end{figure}

In addition to generating mock datasets with statistical fluctuations, we also generate the Asimov dataset \cite{Cowan:2010js}. The Asimov dataset consists of data that are identical to the prediction under the hypothesis being tested, with no statistical fluctuations, even if this means using fractional counts in cases where the data consists of binned counts. In this case, we generate data that lies exactly on top of the predicted density profile for a given mass model.
As shown in Ref.~\cite{Cowan:2010js}, the mean likelihood profile over an ensemble of simulated datasets converges to the likelihood profile obtained with the Asimov dataset in the vicinity of the point of maximum likelihood.
This provides a way of cross-checking both the likelihood function and the mock-data generation framework.

In Fig.~\ref{fig:smear}, we show the likelihood profiles for a sample of mock datasets, as compared with the likelihood profiles computed on the Asimov dataset. When the datasets are generated without scatter due to parallax uncertainty and compared to models without Gaussian kernel smearing, the resulting likelihood profiles behave as expected from the Asimov dataset (left panel). However, when datasets are generated {\emph{with}} scatter from parallax uncertainty and compared to models \emph{without} Gaussian kernel smoothing, we obtain artificially steep likelihood profiles (center panel). Finally, we analyze mock datasets with parallax uncertainty and Gaussian kernel smoothing (right panel), finding agreement with the expectation from the Asimov dataset.
These results indicate the importance of including the Gaussian kernel smoothing for obtaining the correct likelihood profiles, and this also verifies that we have adequately accounted for these effects in our modeling.

\FloatBarrier
\subsection{Results of Mock Data Analysis}

\begin{figure}[t!]
\includegraphics[width = \textwidth]{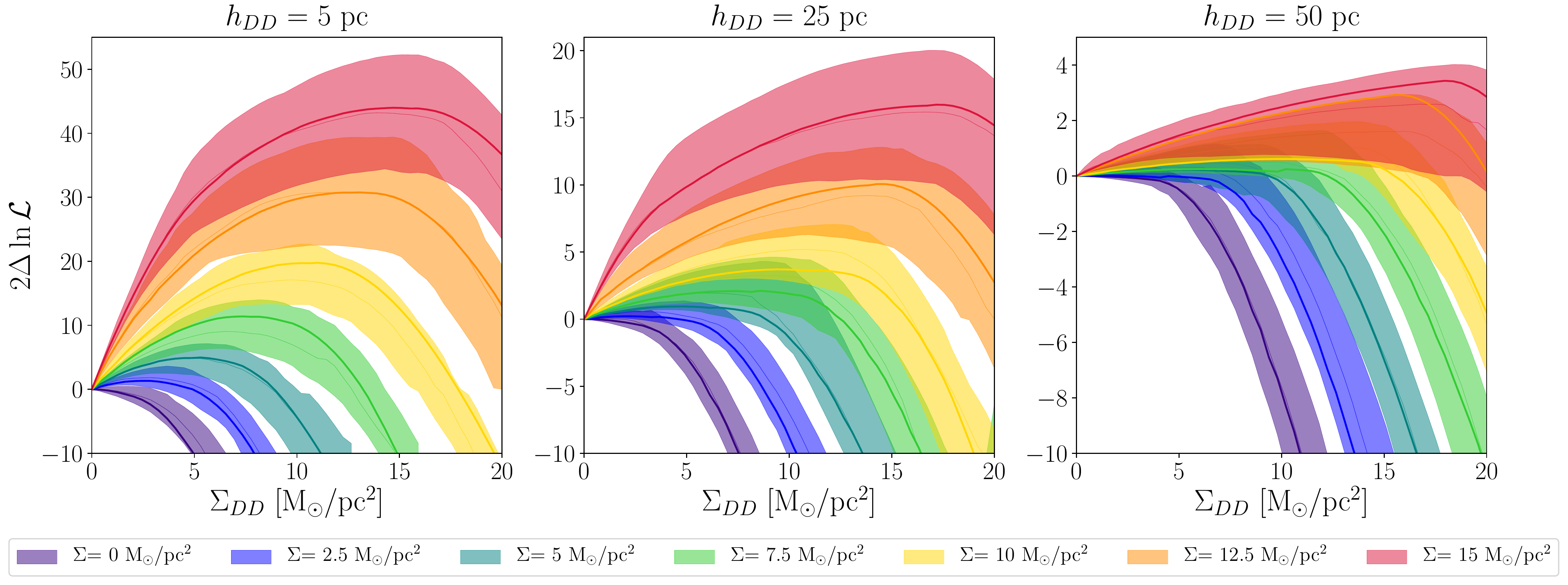}
\caption{The likelihood profiles $2\Delta \ln \mathcal{L}$ for mock data with injected DD signals. Different colors correspond to different values of the injected DD surface density. The bands show the 68\% containment regions for the 50 mock datasets for each set of parameters. The thick line is the likelihood profile of the Asimov dataset, while the thin line is the median likelihood profile of the 50 mock datasets.}
\label{fig:rainbow}
\end{figure}

Using the methods outlined above, we generate mock datasets with DDs of varying surface density between 0 and 15 $\msol$/pc$^2$ and scale heights $h_{DD} = 5, 25$, and 50 pc. For each set of DD parameters, we generate 50 such datasets including the effects of parallax uncertainties and run these datasets through our analysis with Gaussian kernel smoothing. In Fig.~\ref{fig:rainbow}, for each value of $h_{DD}$ we show the resulting likelihood profiles in $\Sigma_{DD}$.
Thinner DDs with larger surface density (and corresponding higher midplane density) are detected with much higher significance using our analysis. For the thickest DD we inject into the data, even large surface density DDs are not detected with high significance, in part because the density $\sim \Sigma_{DD}/4 h_{DD}$ is lower and also because there is a possible degeneracy of the DD with the baryonic disk and halo DM. 

\begin{figure}[t!]
\includegraphics[width = \textwidth]{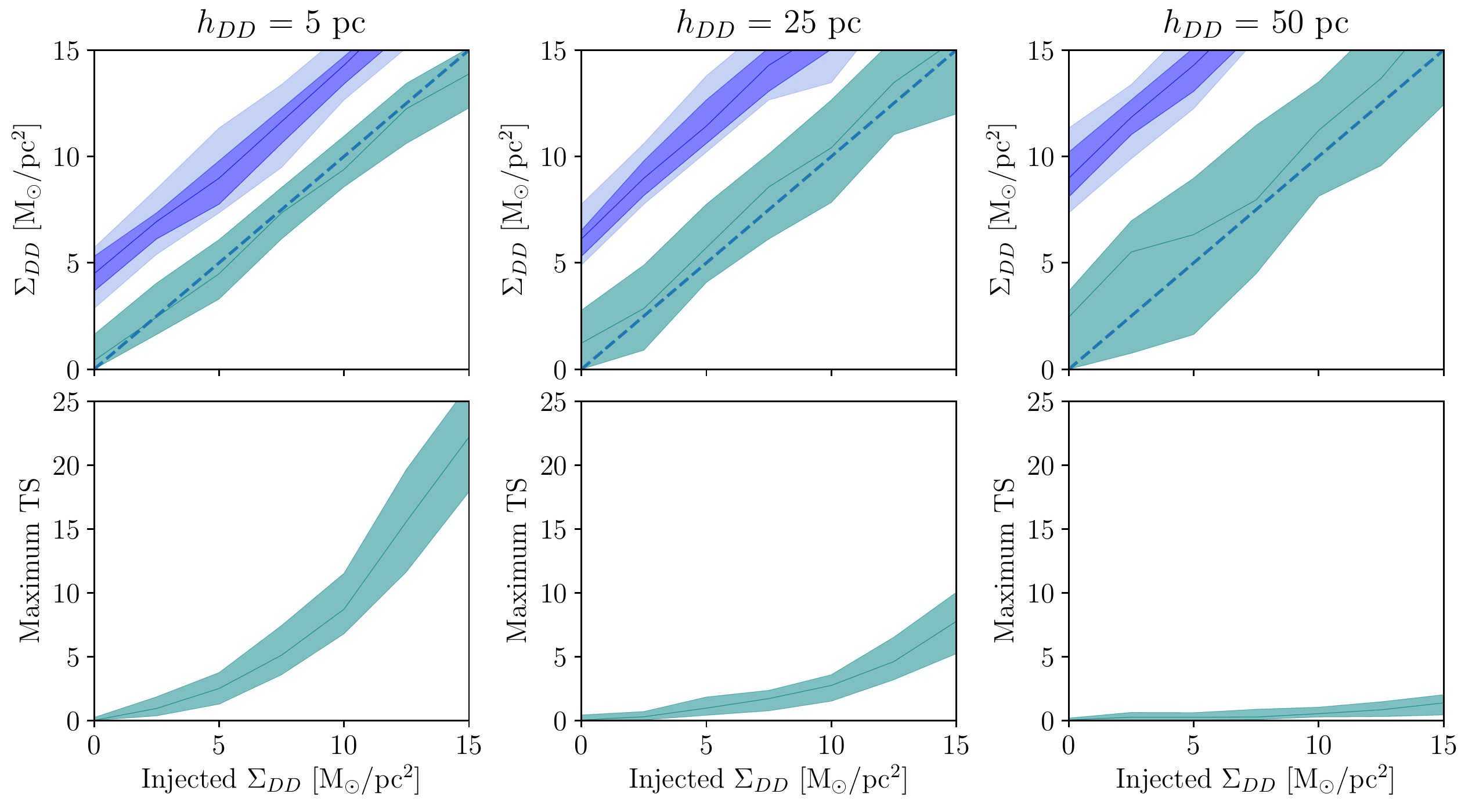}
\caption{Top: For mock data with injected DD signals, the recovered value of $\Sigma_{DD}$ (green band, at 68\% containment) and the 95\% upper limit on $\Sigma_{DD}$ as determined by $2 \Delta \ln \mathcal{L} = -2.71$ (blue bands). The dark (light) blue bands represent the 68\% (95\%) containment regions for the upper limits determined by 50 mock datasets. Crucially, we never exclude the injected signal, which is indicated by the dashed diagonal line. Bottom: The 68\% containment region for the maximum TS obtained in our analysis, which gives a measure of the detection strength of the recovered signal. }
\label{fig:injected}
\end{figure}

The top panel of Fig.~\ref{fig:injected} compares the recovered best fit $\Sigma_{DD}$ (green band) with the injected value (dashed line), where we find good agreement. We also show the 95\% limit obtained from the data sets with injected signals, indicated by the blue band. As desired, we never rule out an injected signal. The fact that even the 95\% containment region of the limit (light blue) does not have any overlap with the injected signal (dashed line) means that our ``95\%'' exclusion --- which is determined by assuming that the likelihood profile is $\chi^2$-distributed in the vicinity of the point of maximum likelihood --- is likely conservative. 

\FloatBarrier

\section{Extended results}
\label{sec:extended}

In this section, we provide details of our main results: we present full likelihood profiles and show the dependence of the final limit on various elements of the analysis. We also give extended results on parameter estimation with a Bayesian framework.

\begin{figure}[t]
\includegraphics[width = 0.5\textwidth]{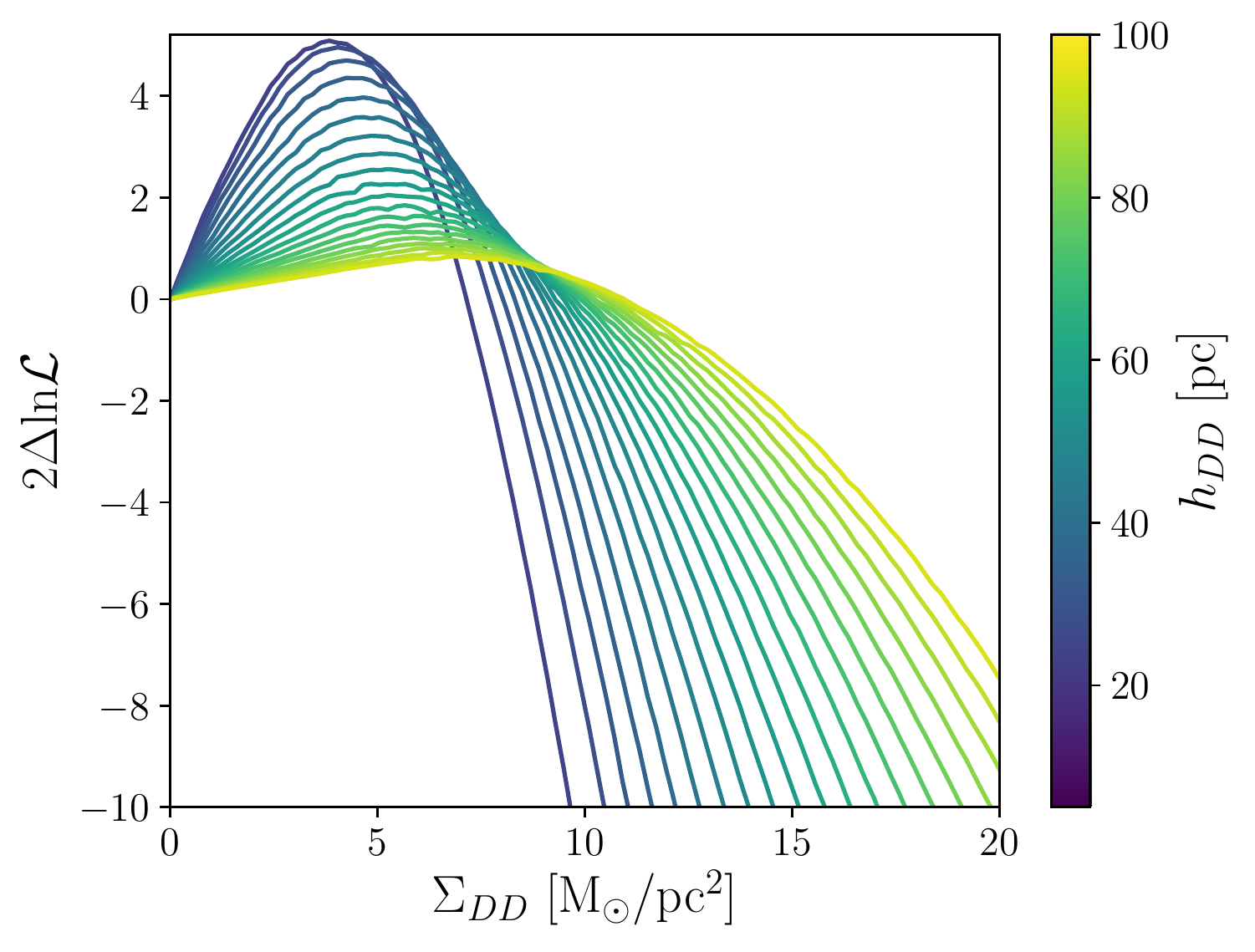}\includegraphics[width = 0.5\textwidth]{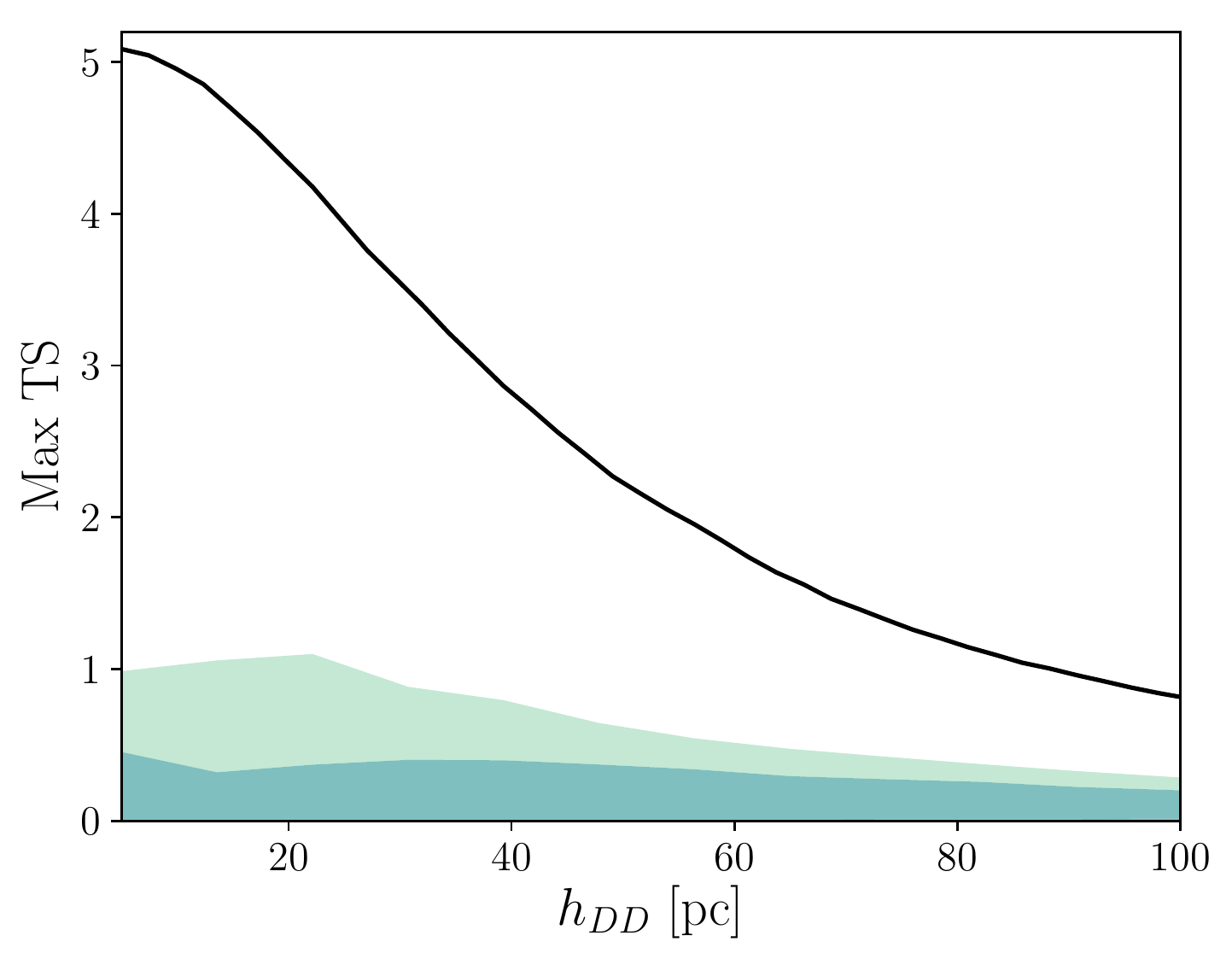}
\caption{Left: For each $h_{DD}$, we show the likelihood profile used to set 95\% one-sided bounds on $\Sigma_{DD}$ in our fiducial analysis. Right: For the same analysis, we show the maximum TS as a function of $h_{DD}$ (black line). The dark and light green regions are 68\% and 95\% containment regions for simulated data generated under the null hypothesis of no DD.  }
\label{fig:LLprofiles}
\end{figure}

For the limit given in Fig.~\ref{fig:exclusion}, we also show the complete likelihood profiles over the range $h_{DD} = 5-100$ pc in Fig.~\ref{fig:LLprofiles} (left panel), with maximum TS $\sim 5$ for the smallest DD scale height in our analysis, $h_{DD} = 5$ pc. The right panel shows the maximum TS as a function of $h_{DD}$, compared to the expectation from 100 mock datasets generated with no DD, similar to the band shown in Fig.~\ref{fig:exclusion}. Although our TS value is well outside of the 95\% containment region of the mock data (light green band), it is not easy to interpret the statistical significance of this result due to the possibility of hidden systematics. If the fiducial model is adjusted to have larger baryonic densities or larger halo DM density, then the expected band of TS values will also increase compared to that shown. It is also possible that the assumptions going into our vertical Poisson-Jeans modeling do not fully capture the dynamics.

\begin{figure}[t]
\includegraphics[width = 0.5\textwidth]{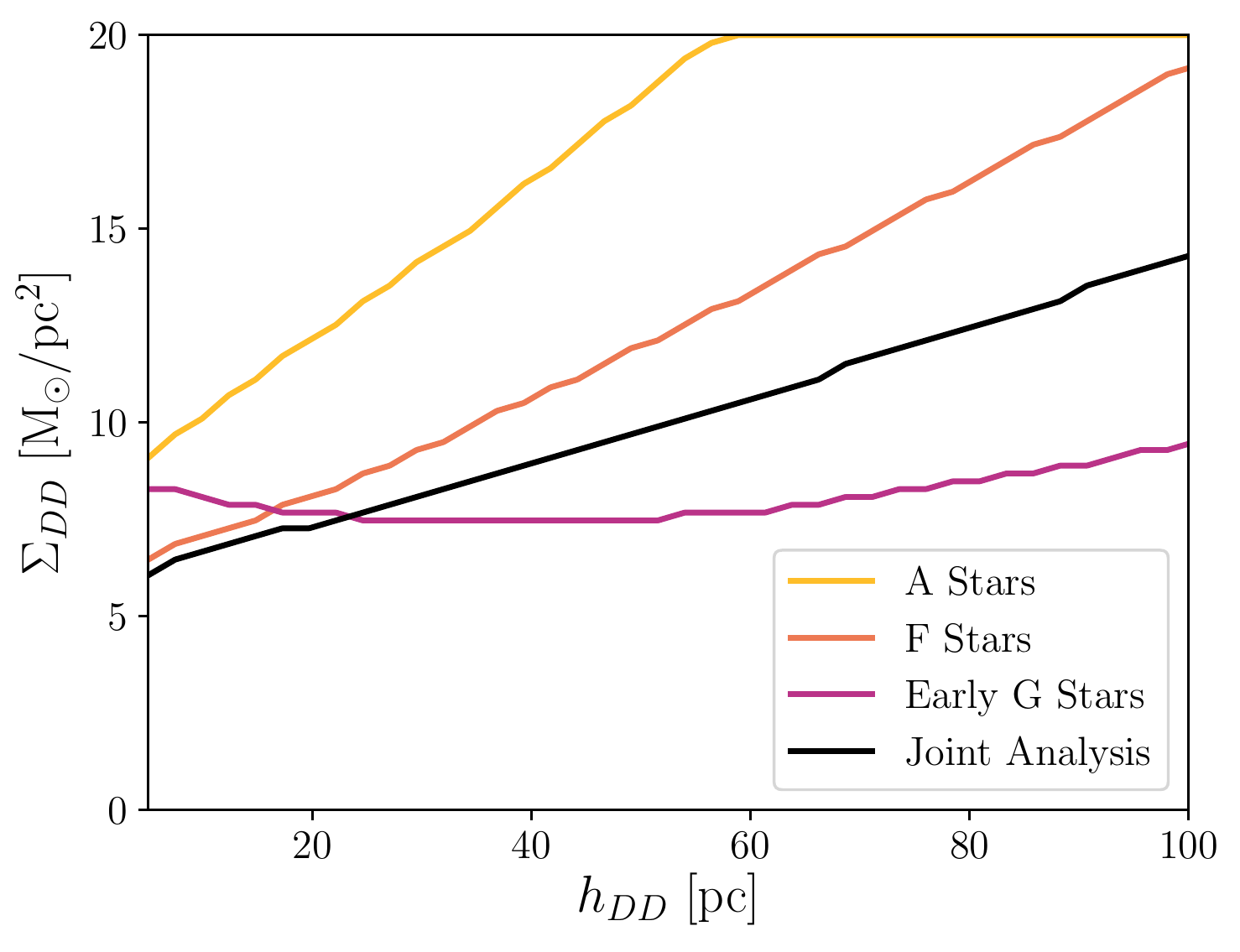}\includegraphics[width = 0.49\textwidth]{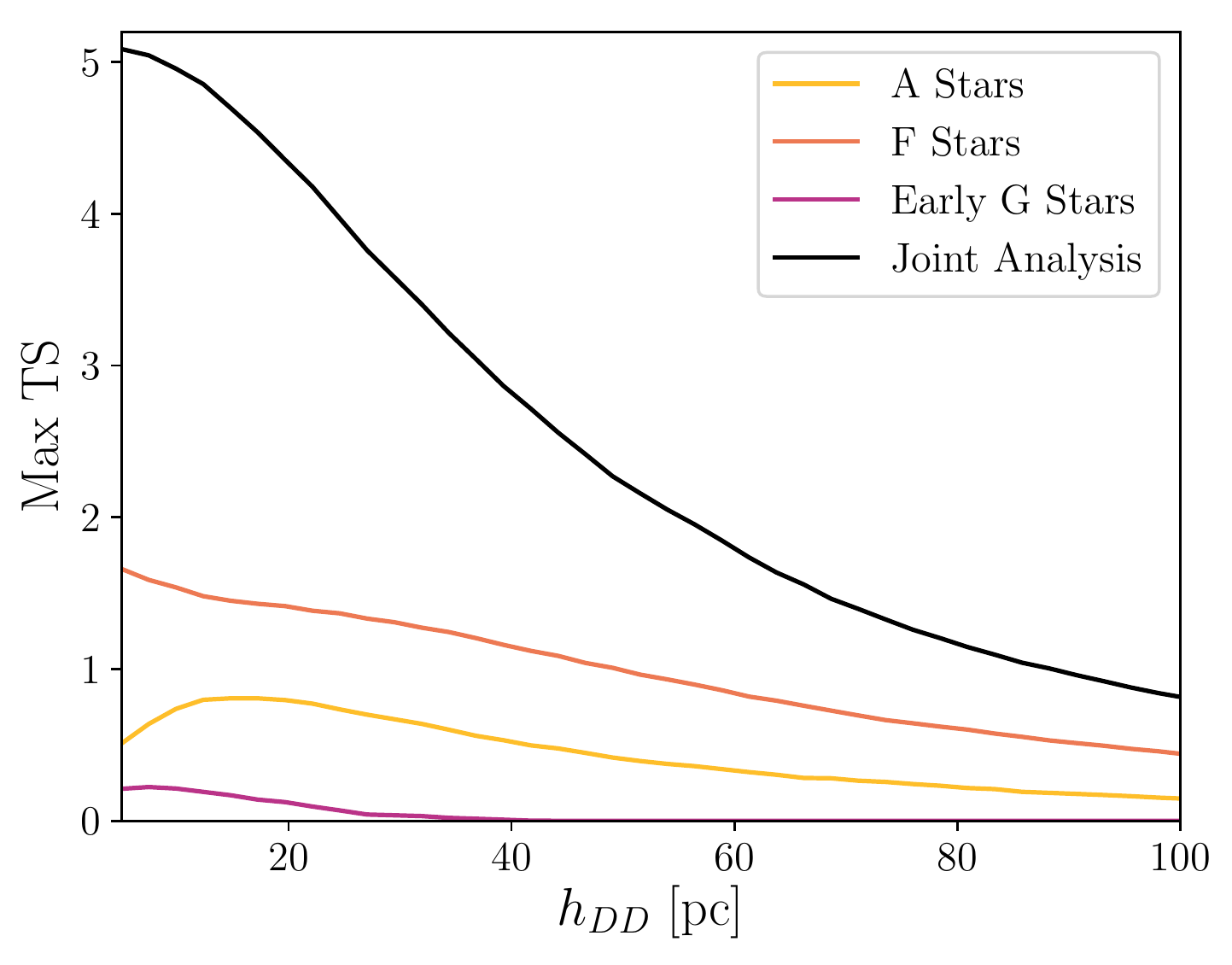}
\caption{Left: The joint analysis on all three tracer populations is compared with the limit obtained from separate analyses on the individual populations. While the limits from only A stars are noticeably weaker, there are roughly ten times fewer stars than either F or Early G stars. Right:  The maximum TS as a function of $h_{DD}$ for each tracer population as well as the joint analysis. 
} 
\label{fig:subtyperesults}
\end{figure}

The joint analysis including A, F, and early G stars is compared to the limits separately obtained on the individual tracer populations in the left panel of Fig.~\ref{fig:subtyperesults}. Here we separately marginalize over the height of the Sun, $\rhoDM$ and baryons for each tracer population, Eq.~\ref{likelihood}. Due to the factor of $\sim$10 fewer stars in the A star sample, our limits are mainly driven by the F and early G stars. The F and early G stars have similar statistics and systematic errors, however the F star data favors a higher DD density and the limits are overall weaker. The right panel of Fig.~\ref{fig:subtyperesults} shows the TS values obtained from the separate analyses, which are lower than that from the joint analysis since the joint result requires the baryon parameters are shared for the three populations. 

\subsection{Dependence of Results on Binning and Gas Densities}

\begin{figure}[t]
\includegraphics[width = \textwidth]{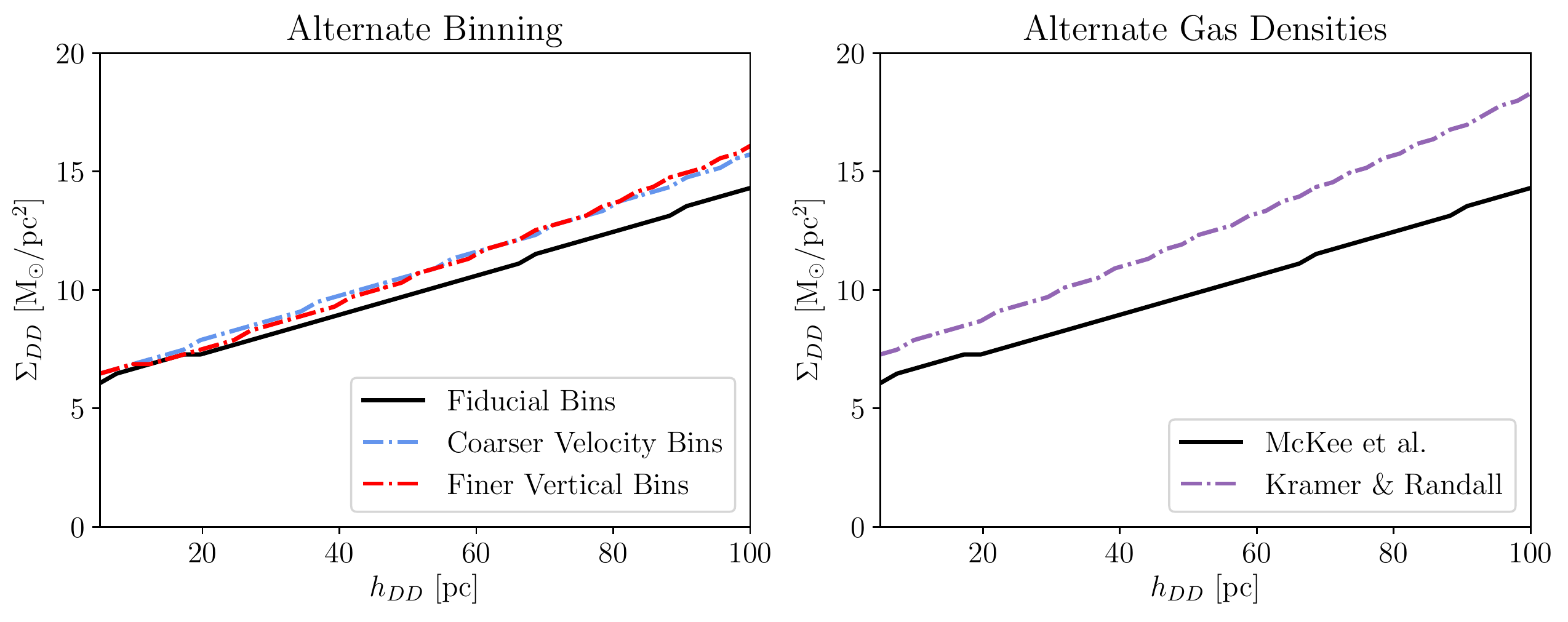}
\caption{Left: Dependence of the limit on the choice of bin sizes. While for the fiducial analysis (black line), we use $z$-bins of width 25 pc and $v$-bins of width 2 km/s, there is a mild difference in the limits when using finer $z$-bins of width 8 pc (red dashed) or when using coarser $v$-bins of width 4 km/s (blue dashed). Right: Dependence of the limit on different models for the gas densities, given in Table~\ref{tab:KR}. Our main analysis uses gas densities from Ref.~\cite{McKee:2015hwa} (McKee et al.), but we also consider the lower gas densities obtained in Ref.~\cite{Kramer:2016dew} (Kramer \& Randall).}
\label{fig:binresults}
\end{figure}

As discussed above, our choice of binning for the data was motivated by the size of typical measurement uncertainties in $z$ and vertical velocities $v$. In Fig.~\ref{fig:binresults} (left panel), we show how the limits change with alternate binning choices, finding our conclusions are robust to these differences. Although the limits are somewhat weaker at large $h_{DD}$ with different $z$-bin or $v$-bin widths, the bounds for a thin DD at $h_{DD} \sim 10$ pc are unchanged.

The right panel of Fig.~\ref{fig:binresults} gives limits under the assumption of a different mass model. Our main mass model in Table~\ref{tab:massmodel} drew on gas densities obtained in Ref.~\cite{McKee:2015hwa} (McKee et al.), while the revised estimates of Ref.~\cite{Kramer:2016dew} (Kramer \& Randall) found lower gas densities. The comparison is given in Table~\ref{tab:KR}. With the Kramer \& Randall gas densities, a larger DD density is allowed and we find moderately weaker limits; however, the 95\% bound still excludes surface densities of $\Sigma_{DD} \sim 10$ M$_\odot$/pc$^2$ for a thin DD with $h_{DD} \sim 10$ pc.

\begin{table}[h]
\begin{center}
\begin{tabular}{l | c c } 
 & \multicolumn{1}{ c}{Kramer \& Randall} & \multicolumn{1}{c }{McKee et al.} \\
Gas Component & \multicolumn{1}{ c}{$\quad \rho(0)$ [M$_\odot$/pc$^3$]} & \multicolumn{1}{c }{$\quad \rho(0)$ [M$_\odot$/pc$^3$] $\quad$} \\
\hline
Molecular Gas (H$_2$) & 0.014$\,\pm\,$0.005 & 0.01$\,\pm\,$0.003 \\
Cold Atomic Gas (H$_\text{I}$(1)) & 0.015$\,\pm\,$0.003 & 0.028$\,\pm\,$0.006 \\
Warm Atomic Gas (H$_\text{I}$(2)) & 0.005$\,\pm\,$0.001 & 0.007$\,\pm\,$0.001  \\
Hot Ionized Gas (H$_\text{II}$) & 0.0011$\,\pm\,$0.0003 & 0.0005$\,\pm\,$0.00002 \\
\end{tabular}
\caption{Midplane gas densities from the Kramer \& Randall~\cite{Kramer:2016dew} compared with those from McKee et al.~\cite{McKee:2015hwa}. \label{tab:KR}}
\end{center}
\vspace{-0.5cm}
\end{table}

\FloatBarrier
\subsection{Bayesian analysis}

Using the nested sampling code \texttt{Multinest}~\cite{Buchner:2014nha}, we have separately analyzed the individual tracer populations to obtain posteriors on the DD and nuisance parameters. This allows us to check consistency of results between the tracers, while differences could be sensitive to nonequilibrium effects as discussed in the main Letter. 

\begin{figure}[h!]
\includegraphics[width = \textwidth]{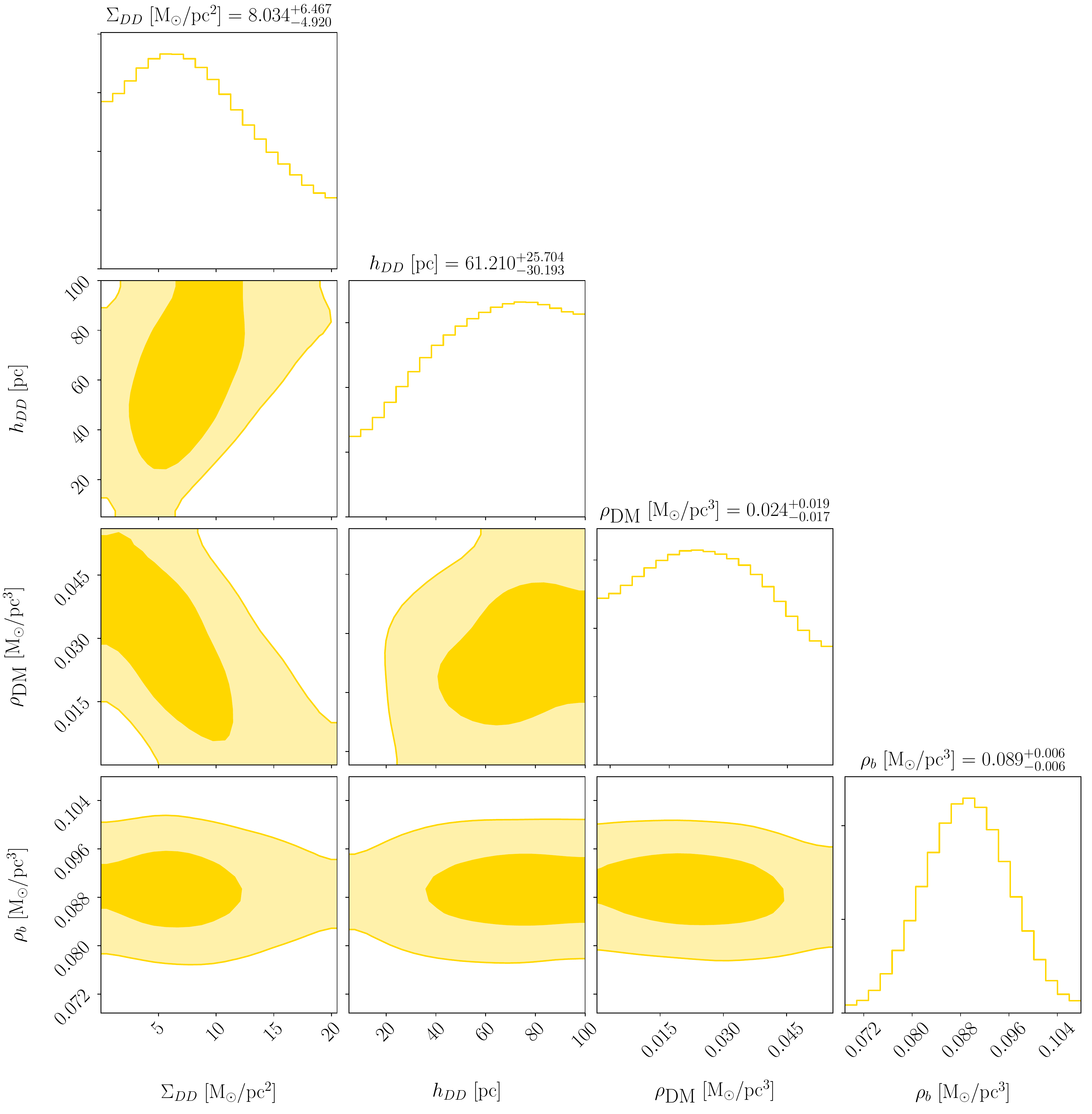}
\caption{ \label{fig:Acorner} Posterior distributions in DD parameters, the local density of the DM halo $\rhoDM$, and the total baryon density $\rho_b$ for an analysis including only the A stars. The dark (light) regions correspond to 1$\,\sigma$ (2$\,\sigma$) containment regions. }
\end{figure}

\begin{figure}[h!]
\includegraphics[width = \textwidth]{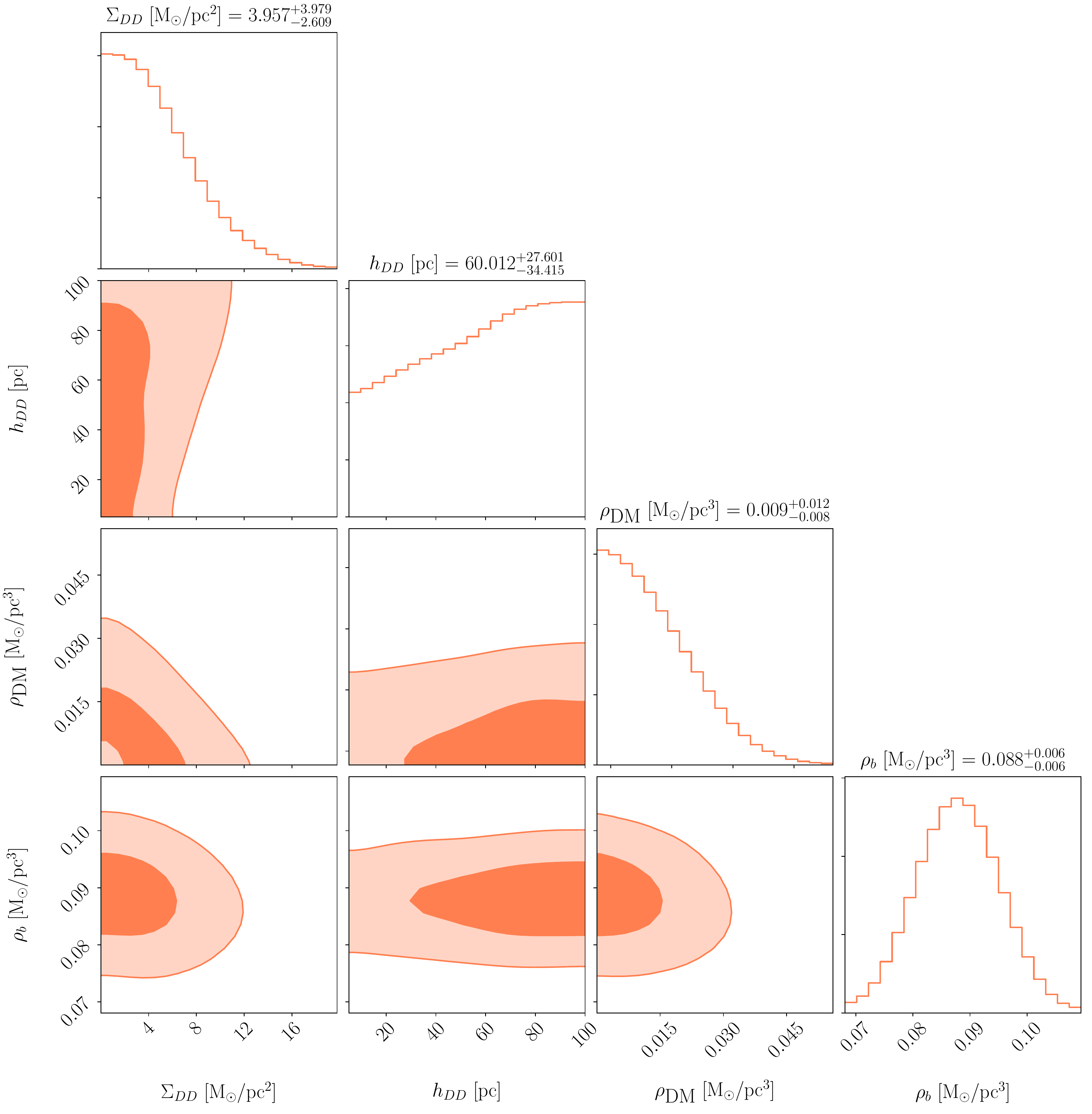}
\caption{  \label{fig:Fcorner} Same as Fig.~\ref{fig:Acorner} except for the F stars.} 
\end{figure}

\begin{figure}[h!]
\includegraphics[width = \textwidth]{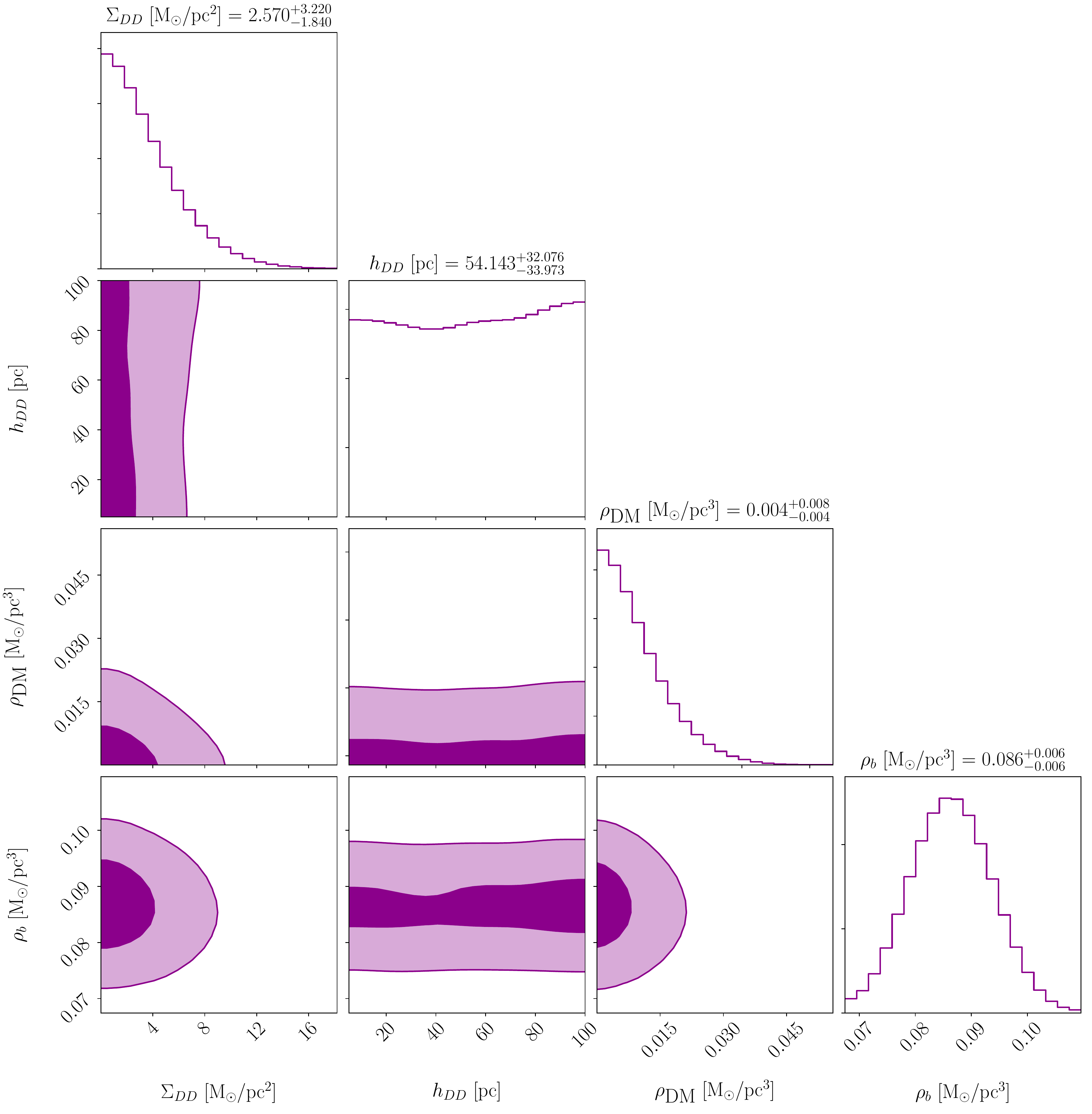}
\caption{  \label{fig:Gcorner} Same as Fig.~\ref{fig:Acorner} except for the early G stars.}
\end{figure}

\begin{figure}[h!]
\includegraphics[width = \textwidth]{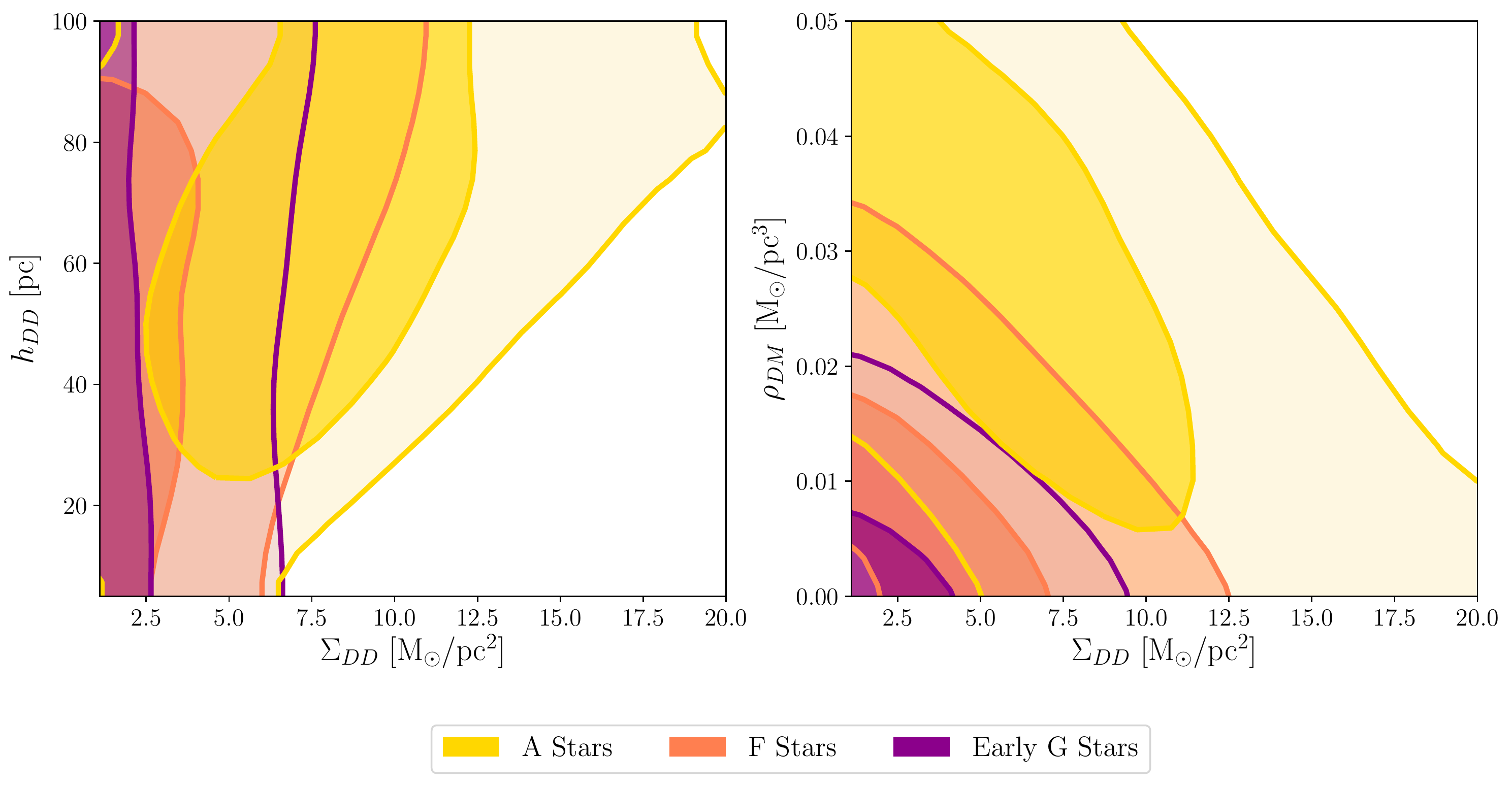}
\vspace{-0.3cm}
\caption{ Here we overlay the posteriors for the DD parameters and halo DM $\rhoDM$, shown in Figs.~\ref{fig:Acorner}-\ref{fig:Gcorner}, for the three tracer populations. The shaded contours are the $1\,\sigma$ (darker) and $2\,\sigma$ (lighter) containment regions.
}
\label{fig:covariance}
\end{figure}

Including all nuisance parameters and a thin DD, we find the posterior distributions shown in Figs.~\ref{fig:Acorner}-\ref{fig:Gcorner} for the DD parameters, halo DM $\rhoDM$, and the total baryon density $\rho_b$. There is a degeneracy between halo DM and thin disk DM, as well as a broad posterior distribution for $h_{DD}$, such that the $z$-dependence of the DM profile is poorly determined near the midplane. 
In Fig.~\ref{fig:covariance}, we overlay the results for the DD parameters and $\rhoDM$ for the three populations, finding that they agree within the $\sim$1-2$\,\sigma$ containment regions. The results in a Bayesian framework are consistent with the limits obtained with the profile likelihood method: for a thin dark disk $h_{DD} \approx 10$ pc, we see that the maximum DD surface density within the $2\sigma$ posterior distribution is $\Sigma_{DD} \approx 6-7$ M$_\odot$/pc$^2$ for all tracer populations. We also find consistency in the height of the sun above the Galactic plane for all three tracer populations: for the A, F, and early G stars we find $z_\text{sun} =-4.0^{+2.4}_{-2.6}$, $1.8^{+1.9}_{-2.0}$, and $-1.6^{+3.7}_{-3.7}$~pc, respectively.


\begin{figure}[h!]
\includegraphics[width = 0.5\textwidth]{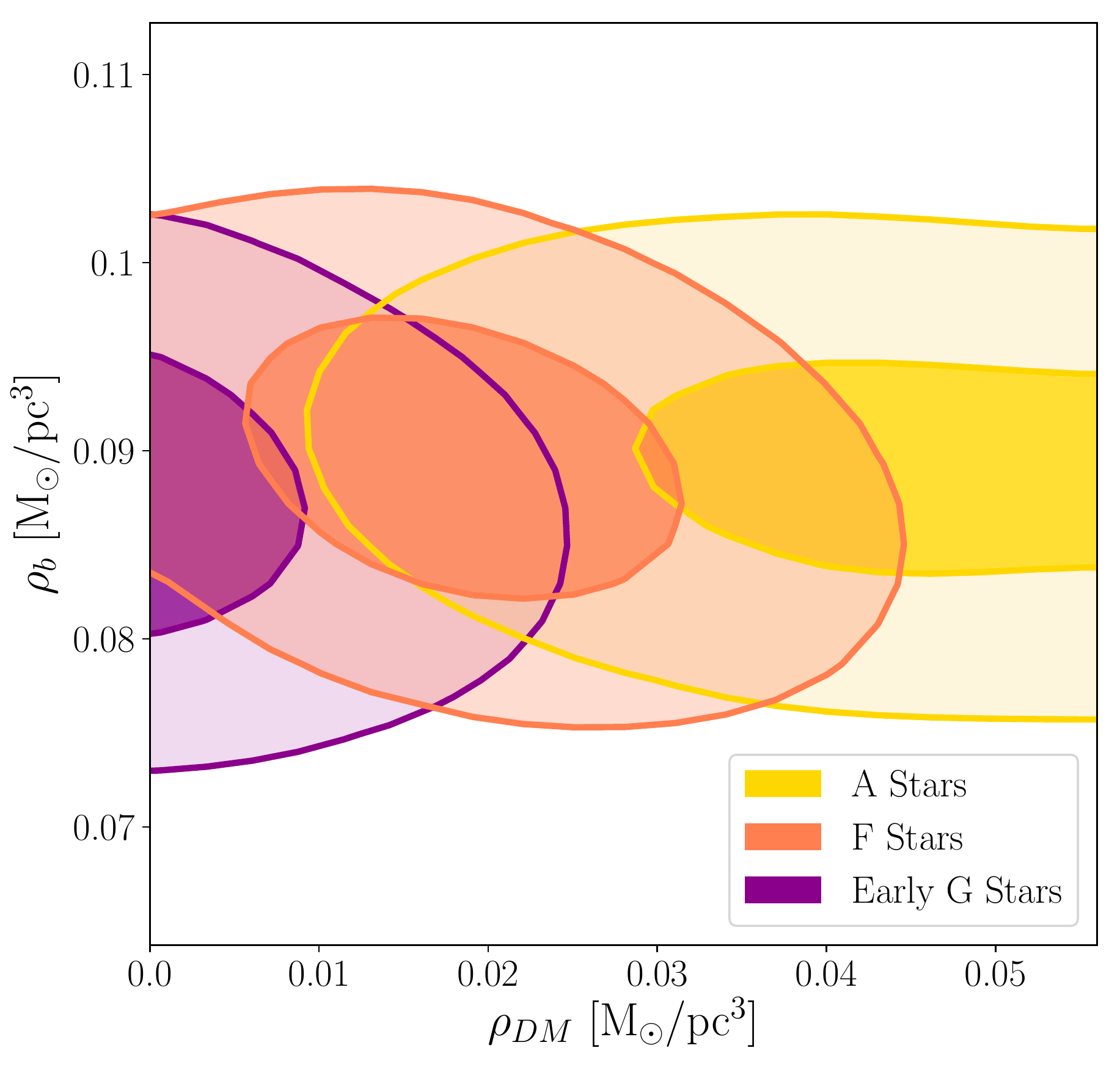}
\vspace{-0.3cm}
\caption{ Assuming no DD is present, marginalized posteriors for the midplane baryon density $\rho_b$ and halo DM $\rhoDM$ for each of the tracer populations. The shaded contours are the $1\,\sigma$ (darker) and $2\,\sigma$ (lighter) containment regions. The complete posteriors for the individual baryon components is given in Table~\ref{tab:G}, while the posterior on $\rho_b$ and $\rhoDM$ is reported in Table~\ref{tab:posterior}.
}
\label{fig:rhob_vs_rhoDM}
\end{figure}

In Fig.~\ref{fig:rhob_vs_rhoDM} we show posterior distributions in $\rhoDM$ and $\rho_b$ for an analysis where no DD is included, which gives a measure of the total midplane density under the null hypothesis. The corresponding posteriors on $\rhoDM$ and $\rho_b$ are given in Table.~\ref{tab:posterior}. Again, we find consistency between the three populations for the $\sim$1-2$\,\sigma$ containment regions. The complete baryon parameters in this case are given in Table~\ref{tab:G}. We have additionally used these to compute the posterior on the baryon surface density in our sample volume as $\Sigma = 25.4^{+1.6}_{-1.6}$, $26.0^{+1.9}_{-1.9}$, and $26.1^{+1.6}_{-1.5}~\msol/$pc$^2$ for the A, F, and G stars, respectively. We can also extrapolate our baryon density profiles out to $z=1.1$~kpc to compute $\Sigma_{1.1}$, a metric often quoted in the literature. In doing this, we extend the assumption of isothermal mass components to beyond the regime for our analysis, where it is expected that velocity dispersions are $z$-dependent. Nevertheless, we find surface densities that agree well with those in the literature. For the A, F, and early G stars we find $\Sigma_{1.1} = 40.7^{+4.0}_{-4.9}$, $44.6^{+4.9}_{-5.5}$, and $48.5^{+5.0}_{-4.8}~\msol/$pc$^2$, respectively. The larger errors at large $z$ reflect the fact that our extrapolation is  sensitive to small variations in the profile within $z<200$~pc.

\subsection{Stability of a thin dark disk}

\rev{Here we briefly comment on the stability of the thin DDs constrained by our analysis. As a basic check, we apply Toomre's criterion to the thin DD in two limiting cases: a collisionless gas, or a cold fluid with speed of sound $c_s$. 

For the case of a self-gravitating collisionless gas, the Toomre stability parameter is  \beq Q = {\sigma_R \kappa}/({3.36 \,G\, \Sigma_{DD}}),\eeq
where $\sigma_R$ is the radial velocity dispersion and $\kappa$ is the epicycle frequency. For a self-gravitating isothermal disk, the vertical velocity dispersion is related to the height and surface density as \beq \sigma_z = \sqrt{2 \pi G \,\Sigma_{DD} h_{DD}}.\eeq Assuming that $\sigma_R \simeq 2 \sigma_z$ as for the stars in the MW, then the condition on stability can be written in terms of dark disk parameters as \beq Q \simeq \frac{2 \sqrt{2 \pi} \kappa}{3.36} \sqrt{\frac{h_{DD}}{G\, \Sigma_{DD}}} > 1.\eeq Taking measured values of $\kappa$~\cite{binney2011galactic}, we find that all the DDs we constrain are stable (except the thinnest DDs with $h_{DD} \lesssim 7$ pc) in this case.

In the latter case of a cold fluid, the  Toomre stability parameter is \beq Q =  {c_s \kappa}/({ \pi G \,\Sigma_{DD}}).\eeq For a self-gravitating fluid in hydrostatic equilibrium, if we assume that the speed of sound is related to the scale height as $c_s = \sqrt{2 \pi G \Sigma_{DD} h_{DD}}$ similar to the collisionless case, then we find that the DD is less stable; however, the DDs we constrain with $h_{DD} \gtrsim 13$ pc are within the regime of stability.  For an isothermal fluid rotating about a central mass, the scale height is instead given by $h_{DD} = {c_s}/({\sqrt{2} \Omega})$ where $\Omega$ is the rotational angular velocity of circular orbits at the Sun's Galactocentric radius. The condition for stability can then be written as \beq Q = {\sqrt{2} h_{DD} \Omega\, \kappa}/({ \pi G \,\Sigma_{DD}} )\gtrsim 1.\eeq For $h_{DD} = 10$ pc, this implies that only dark disks of $\Sigma_{DD} \lesssim 1-2 M_\odot/$pc$^2$ are stable, in contrast to the other scenarios. As demonstrated, the stability of the thin DD depends sensitively on the dynamics leading to DD formation and the properties of the dark sector. In addition, the presence of additional mass components could change these estimates substantially by adding a stabilizing potential \cite{Jog:2013zra}.}

\begin{table}[h]
\begin{center}
\begin{tabular}{l | c c | c c | c c} 
 & \multicolumn{2}{c}{A Stars} &\multicolumn{2}{c}{F Stars}& \multicolumn{2}{c}{Early G Stars} \\ 
Baryonic Component & \multicolumn{1}{ c}{$ \rho(0)$ [M$_\odot$/pc$^3$]} & \multicolumn{1}{c }{$\quad \sigma $ [km/s] $\quad$}& \multicolumn{1}{ c}{$ \rho(0)$ [M$_\odot$/pc$^3$]} & \multicolumn{1}{c }{$\quad \sigma $ [km/s] $\quad$}& \multicolumn{1}{ c}{$ \rho(0)$ [M$_\odot$/pc$^3$]} & \multicolumn{1}{c }{$\quad \sigma $ [km/s] $\quad$} \\
\hline
Molecular Gas (H$_2$) & $0.0105^{+0.0028}_{-0.0028}$&$3.7^{+0.2}_{-0.2}$&$0.0106^{+0.0029}_{-0.0029}$&$3.7^{+0.2}_{-0.2}$&$0.0103^{+0.003}_{-0.0029}$&$3.7^{+0.2}_{-0.2}$ \\
Cold Atomic Gas (H$_\text{I}$(1)) & $0.0283^{+0.0047}_{-0.005}$&$7.1^{+0.4}_{-0.4}$&$0.028^{+0.0051}_{-0.0051}$&$7.1^{+0.5}_{-0.5}$&$0.0265^{+0.0051}_{-0.0049}$&$7.1^{+0.5}_{-0.5}$ \\
Warm Atomic Gas (H$_\text{I}$(2)) & $0.0073^{+0.0006}_{-0.0006}$&$22.0^{+2.3}_{-2.1}$&$0.0073^{+0.0006}_{-0.0007}$&$22.1^{+2.2}_{-2.2}$&$0.0073^{+0.0007}_{-0.0007}$&$22.1^{+2.3}_{-2.2}$ \\
Hot Ionized Gas (H$_\text{II}$) & $0.0005^{+0.00002}_{-0.00002}$&$39.1^{+3.6}_{-3.6}$&$0.0005^{+0.00002}_{-0.00002}$&$39.0^{+3.7}_{-3.8}$&$0.0005^{+0.00002}_{-0.00002}$&$39.0^{+3.8}_{-3.8}$ \\
Giant Stars & $0.0006^{+0.0001}_{-0.0001}$&$15.5^{+1.5}_{-1.4}$&$0.0006^{+0.0001}_{-0.0001}$&$15.5^{+1.5}_{-1.5}$&$0.0006^{+0.0001}_{-0.0001}$&$15.5^{+1.5}_{-1.5}$ \\
$M_V <3$ & $0.0018^{+0.0002}_{-0.0002}$&$7.6^{+1.8}_{-1.9}$&$0.0018^{+0.0002}_{-0.0002}$&$7.5^{+1.9}_{-1.8}$&$0.0018^{+0.0002}_{-0.0002}$&$7.5^{+1.8}_{-1.9}$ \\
$3<M_V<4$ & $0.0018^{+0.0002}_{-0.0002}$&$12.1^{+2.2}_{-2.2}$&$0.0018^{+0.0002}_{-0.0002}$&$12.0^{+2.2}_{-2.2}$&$0.0018^{+0.0002}_{-0.0002}$&$12.0^{+2.2}_{-2.3}$ \\
$4<M_V<5$ & $0.0029^{+0.0003}_{-0.0003}$&$18.0^{+1.6}_{-1.7}$&$0.0029^{+0.0003}_{-0.0003}$&$18.0^{+1.7}_{-1.7}$&$0.0029^{+0.0003}_{-0.0003}$&$18.0^{+1.7}_{-1.7}$ \\
$5<M_V<8$ & $0.0072^{+0.0007}_{-0.0007}$&$18.5^{+1.7}_{-1.7}$&$0.0072^{+0.0007}_{-0.0007}$&$18.5^{+1.8}_{-1.8}$&$0.0072^{+0.0007}_{-0.0007}$&$18.5^{+1.8}_{-1.8}$ \\
$M_V>8$ (M Dwarfs) & $0.0218^{+0.0025}_{-0.0026}$&$18.5^{+3.6}_{-3.5}$&$0.0215^{+0.0026}_{-0.0025}$&$18.4^{+3.7}_{-3.8}$&$0.0211^{+0.0026}_{-0.0025}$&$18.3^{+3.8}_{-3.8}$ \\
White Dwarfs & $0.0056^{+0.0009}_{-0.0009}$&$20.1^{+4.6}_{-4.4}$&$0.0056^{+0.0009}_{-0.001}$&$19.9^{+4.6}_{-4.6}$&$0.0055^{+0.0009}_{-0.0009}$&$19.9^{+4.7}_{-4.6}$ \\
Brown Dwarfs & $0.0015^{+0.0005}_{-0.0004}$&$19.9^{+4.5}_{-4.3}$&$0.0015^{+0.0005}_{-0.0005}$&$19.9^{+4.6}_{-4.7}$&$0.0015^{+0.0005}_{-0.0005}$&$20.0^{+4.7}_{-4.6}$ \\
\end{tabular}
\caption{Marginalized posteriors of baryon parameters for A, F, and early G stars assuming no DD is present. 
\label{tab:G}}
\end{center}
\vspace{-0.5cm}
\end{table}

\end{document}